\documentclass[preprint,times]{elsarticle}
\usepackage[headsep=.2in,footskip=.2in]{geometry}
\usepackage[bookmarks=false]{hyperref} 
\hypersetup{
	colorlinks=true,
	linkcolor=blue,
	filecolor=magenta,      
	urlcolor=cyan,
	citecolor=blue,
}
\usepackage{booktabs}
\usepackage{titlesec} 

\makeatletter
\def\ps@pprintTitle{%
	\let\@oddhead\@empty
	\let\@evenhead\@empty
	\let\@oddfoot\@empty
	\let\@evenfoot\@oddfoot}
\makeatother

\usepackage{amsmath}
\usepackage{amssymb}

\biboptions{authoryear}

\usepackage[figuresright]{rotating}

\usepackage[bookmarks=false]{hyperref}
    \hypersetup{colorlinks,
      linkcolor=blue,
      citecolor=blue,
      urlcolor=blue}

\usepackage{booktabs}
\usepackage{titlesec} 

\usepackage{paracol} 
\columnratio{0.6,0.4} 
\usepackage{fancyhdr}
\usepackage{color}

\usepackage{graphicx} 
\graphicspath{ {./} }
\usepackage[font=normalsize,skip=0pt]{caption}
\usepackage{subcaption}
\usepackage{wrapfig} 
\usepackage{multirow}
\usepackage{array} 

\usepackage{enumitem} 
\usepackage[mathcal]{euscript}

\usepackage[british]{babel}

\usepackage{ifthen} 
\usepackage{lscape} 

\usepackage{pifont} 
\newcommand{\cmark}{\ding{51}}%

\usepackage[usenames,svgnames,table]{xcolor} 
\definecolor{shade}{HTML}{F5DD9D} 

\usepackage{algorithm,algpseudocode}

\newcolumntype{C}[1]{>{\centering\arraybackslash}p{#1}}
\newcolumntype{L}[1]{>{\arraybackslash}p{#1}}
\newcommand{\mrc}[3]{\multirow{#1}{#2}{\centering #3}} 

\definecolor{LightRed}{rgb}{1,0.88,1}
\definecolor{LightGreen}{rgb}{0.88,1,0.88}
\definecolor{Cyan}{rgb}{0.88,1,1}
\definecolor{Black}{rgb}{0.0, 0.0, 0.0}
\definecolor{Blue}{rgb}{0.2, 0.6, 1}
\definecolor{Yellow}{rgb}{1, 1, 0}
\definecolor{DarkYellow}{rgb}{0.922, 0.773, 0.176}
\definecolor{Red}{rgb}{1, 0, 0}
\definecolor{Orange}{rgb}{1, 0.6, 0.2}
\definecolor{Green}{rgb}{0.2,0.8,0.2}
\definecolor{Purple}{rgb}{0.6, 0.4, 1}
\definecolor{Grey}{RGB}{179, 179, 179}

\newsavebox{\verbbox} 

\usepackage{tikz} 
\usepackage{calc} 
\usetikzlibrary{calc,arrows}

\usepackage{acro}
\DeclareAcronym{IATA}{ short = IATA, long = International Air Transport Association, short-plural = s, long-plural = s}
\DeclareAcronym{IOSA}{ short = IOSA, long = IATA Operational Safety Audit, short-plural = s, long-plural = s}
\DeclareAcronym{PASO}{ short = PASO, long = Pacific Aviation Safety Office, short-plural = s, long-plural = s}
\DeclareAcronym{ICAO}{ short = ICAO, long = International Civil Aviation Organisation, short-plural = s, long-plural = s}
\DeclareAcronym{ABM}{ short = ABM, long = agent-based model, short-plural = s, long-plural = s}
\DeclareAcronym{ASM}{ short = ASM, long = available seat mile, short-plural = s, long-plural = s}
\DeclareAcronym{RPM}{ short = RPM, long = passenger revenue mile, short-plural = s, long-plural = s}
\DeclareAcronym{EM}{ short = EM, long = expectation maximization, short-plural = s, long-plural = s}
\DeclareAcronym{GA}{ short = GA, long = genetic algorithm, short-plural = s, long-plural = s}
\DeclareAcronym{PSO}{ short = PSO, long = particle swarm optimization, short-plural = s, long-plural = s}
\DeclareAcronym{MADS}{ short = MADS, long = mesh adaptive direct search, short-plural = s, long-plural = s}
\DeclareAcronym{GPU}{ short = GPU, long = graphics processing unit, short-plural = s, long-plural = s}
\DeclareAcronym{RM}{ short = RM, long = revenue management, short-plural = s, long-plural = s}
\DeclareAcronym{MLE}{ short = MLE, long = maximum likelihood estimation, short-plural = s, long-plural = s}
\DeclareAcronym{QMC}{ short = QMC, long = quasi-Monte Carlo, short-plural = s, long-plural = s}
\DeclareAcronym{RNG}{ short = RNG, long = random number generation, short-plural = s, long-plural = s}
\DeclareAcronym{IPSW}{ short = IPSW, long = industrial problem solving workshop, short-plural = s, long-plural = s}
\DeclareAcronym{CRS}{ short = CRS, long = computer reservation system, short-plural = s, long-plural = s}
\DeclareAcronym{KPI}{ short = KPI, long = key performance indicator, short-plural = s, long-plural = s}
\DeclareAcronym{AAI}{ short = AAI, long = Atmosfair Airline Index, short-plural = s, long-plural = s}
\DeclareAcronym{ATDB}{ short = ATDB, long = AeroTransport Data Bank, short-plural = s, long-plural = s}
\DeclareAcronym{LCC}{ short = LCC, long = low-cost carrier, short-plural = s, long-plural = s}
\DeclareAcronym{ULCC}{ short = ULCC, long = ultra low-cost carrier, short-plural = s, long-plural = s}
\DeclareAcronym{GDP}{ short = GDP, long = gross domestic product, short-plural = s, long-plural = s}
\DeclareAcronym{ER}{ short = ER, long = entity-relationship, short-plural = s, long-plural = s}
\DeclareAcronym{NLP}{ short = NLP, long = natural language processing, short-plural = s, long-plural = s}
\DeclareAcronym{GNN}{ short = GNN, long = graph neural network, short-plural = s, long-plural = s}
\DeclareAcronym{GCN}{ short = GCN, long = graph convolutional network, short-plural = s, long-plural = s}
\DeclareAcronym{GAT}{ short = GAT, long = graph attention network, short-plural = s, long-plural = s}
\DeclareAcronym{MLP}{ short = MLP, long = Multilayer perceptron, short-plural = s, long-plural = s}
\DeclareAcronym{LSTM}{ short = LSTM, long = long short-term memory, short-plural = s, long-plural = s}
\DeclareAcronym{JK}{ short = JK, long = jumping knowledge, short-plural = s, long-plural = s}
\DeclareAcronym{ILP}{ short = ILP, long = integer linear program, short-plural = s, long-plural = s}
\DeclareAcronym{MIQP}{ short = MIQP, long = mixed integer quadratic program, short-plural = s, long-plural = s}
\DeclareAcronym{HHI}{short = HHI, long = Herfindahl-Hirschman Index, short-plural = s, long-plural = s}
\DeclareAcronym{MPC}{short = MPC, long = market penetration capability, short-plural = s, long-plural = s}
\DeclareAcronym{MIP}{short = MIP, long = mixed integer programming, short-plural = s, long-plural = s}
\DeclareAcronym{PMF}{short = PMF, long = probability mass function, short-plural = s, long-plural = s}
\DeclareAcronym{PWL}{short = PWL, long = piecewise linear approximation, short-plural = s, long-plural = s}
\DeclareAcronym{CDF}{short = CDF, long = cumulative distribution function, short-plural = s, long-plural = s}
\DeclareAcronym{CMF}{short = CMF, long = cumulative mass function, short-plural = s, long-plural = s}
\DeclareAcronym{OAG}{short = OAG, long = Official Aviation Guide of the Airways, short-plural = s, long-plural = s}

\DeclareSymbolFont{bbold}{U}{bbold}{m}{n}
\DeclareSymbolFontAlphabet{\mathbbold}{bbold}

\newif\ifnotes\notestrue
\notesfalse             
\def\boxnote#1#2{\ifnotes\fbox{\footnote{\ }}\ \footnotetext{ From #1: #2}\fi}

\def\fabian#1{\boxnote{Fabian}{\color{black}#1}}
\def\mfabian#1{{\color{black} #1}}
\def\hfabian#1{}

\def\mkhalil#1{{\color{black} #1}}
\def\hkhalil#1{}

\usepackage{soul}

\begin{document}
\begin{frontmatter}




\title{Analyzing Airline Alliances through Multi-Attribute Graph Partitioning to Maximize Competition and Market Penetration Capability}


\author[a]{Khalil Al Handawi\corref{cor1}} 
\author[b]{Fabian Bastin}

\address[a]{McGill University, Department of Mechanical Engineering, Macdonald-Stewart Building, 817 Sherbrooke W., Montreal, QC, Canada H3A 2T7}
\address[b]{University of Montreal, Department of Computer Science And Operations Research, CP 6128 Succ Centre-Ville, Montreal, QC, Canada  H3C 3J7}

\begin{abstract}

The air transportation market is highly competitive and dynamic. Airlines often form alliances to expand their network reach, improve operational efficiency, and enhance customer experience. However, the impact of these alliances on market competition and operational efficiency is not fully understood. In this paper, we propose a novel approach to analyze airline alliances using multi\mfabian{-}attribute graph partitioning. We develop metrics to quantify the competitiveness of flight segments and the market penetration capability of airlines based on their alliance memberships. We formulate a bi\mfabian{-}objective optimization problem to maximize both competition and market penetration simultaneously. We also propose algorithms to solve this optimization problem and demonstrate their effectiveness using real-world flight schedule data. Our results provide insights into the structure of airline alliances and their implications for market competition and operational efficiency.

\end{abstract}

\cortext[cor1]{Corresponding author. Tel.: +1-514-572-7367.}
\end{frontmatter}

\email{khalil.alhandawi@mail.mcgill.ca}



\section{Introduction} \label{sec:introduction}

Air transportation involves complex logistics and specifics for transporting passengers around the globe.
Airlines commonly \textit{interline} with \mfabian{competitors}\hfabian{other airlines} to allow their passengers to switch airlines along routes.
This practice allows airlines to offer \hfabian{their passengers} more destinations and expand their network.
Airlines tend to collaborate closely on specific routes and aspects of air transportation to provide an attractive ``joint'' product to passengers that goes beyond interlining. This cooperation is often formalized in the form of:
\begin{itemize}
	\item \textbf{Alliances:} where airlines cooperate on a broader scale, and on offering reciprocal benefits to passengers.
	\item \textbf{Antitrust immunity:} where airlines are allowed to cooperate on pricing and scheduling without violating antitrust laws.
	\item \textbf{Codesharing agreements:} where airlines share the same physical flight but are able to independently sell tickets under their own brand.
	\item \textbf{Joint ventures:} where airlines cooperate on a specific route or set of routes by sharing revenue, costs, and coordinating schedules.
\end{itemize}

Airline alliances and other forms of cooperation also benefit passengers in terms of lower interline fares by internalizing the costs to the airlines. The airlines simultaneously benefit from the increased passenger volumes and the ability to offer a more comprehensive network to passengers \citep{Brueckner2000}. However, as soon as consolidation decreases competition, passengers may lose depending on the market power effect \citep{Bilotkach2019}. While airline alliances are thought to enhance network efficiency and market reach, the actual impacts can be complex and varied. The evidence suggests that not all alliances lead to positive outcomes in terms of increased traffic or competition \citep{OumParkZhan96,Hanl07,Pitfield2007}. 

Theoretical and empirical studies have affirmed that parallel partnerships (i.e., alliances between airlines that compete on the same routes) are anti-competitive and lead to negative market consolidations while complementary partnerships (i.e., alliances between airlines that do not compete on the same routes) are pro-competitive \citep{OumParkKimYu04,Bilotkach2019a}. However, some studies have suggested that this hypothesis does not always hold true and parallel partnerships can yield benefits to consumers even on the overlapping parts of the joint network due to increased volumes and the economies of traffic density~\citep{Brueckner2000}.

Current theoretical research does not fully describe airline partnerships in the context of antitrust immunity (e.g., Open Skies) and joint ventures which are a necessary condition for such partnerships and are often contingent on the approval of government regulators such as the US Department of Justice Antitrust Division. There are a limited number of theoretical and empirical efforts to distinguish between the partnerships that are covered by antitrust immunity and those that are not \citep{Brueckner2003,Bilotkach2005,Bilotkach2011}. However, these studies focus on specific airline partnership structures such as alliances in isolation and do not fully study the interactions between different partnership types and their combined effects on competition and interline fares. Recent studies have attempted to fill this gap by examining the competitive implications of metal-neutral joint ventures and immunized alliances, yet the results remain mixed and market-specific~\citep{CalzEilaIsra17}.

Furthermore, airline partners often cite the increase in their efficiency as part of their antitrust immunity applications~\citep{OumYuZhan01, Pitfield2007}. The most commonly used metric for operational efficiency is the load factor defined as the ratio of passenger miles flown to \acp{ASM}. It is a measure of how efficiently an airline fills its seats. The load factor is a key metric in the airline industry as it is a measure of how well an airline is utilizing its capacity~\citep{DanaGree19}.

Theoretical studies have focused mostly on price effects of airline partnerships. Effects on passenger volumes were automatically realized through the law of demand (i.e., the lower the price, the higher the volume). Some models include economies of traffic density, but they do not explicitly model load factor effects. Such measures are difficult to determine for a regulator due to the proprietary nature of the required data. It is therefore important for regulators to understand the effects of airline partnerships on competition and market consolidation to make informed decisions on the approval of such partnerships~\citep{Brue01}.

Further, a large body of empirical research has focused on US domestic datasets that include at least one flight segment marketed by a US carrier \citep{Bilotkach2019}, as US antitrust laws require airlines to report data on their domestic flights while international flights are not subject to the same reporting requirements. This has led to a lack of empirical studies on international flights and the effects of airline partnerships on international routes \citep{Bilotkach2019a,Brueckner2000}. By looking at international routes, we can gain a better understanding of the effects of airline partnerships on competition and market consolidation on a global scale \citep{OumParkZhan96, Bilotkach2019}.

In this work, we focus on a particular airline partnership structure—the airline alliance membership—since it affects the airline network globally and is not specific to any particular domestic market or route.
We focus on how alliances can be structured more effectively to truly maximize both competition and the capability of airlines to expand their route networks without leading to negative market consolidations.
We first formalize the problem of airline alliance membership as a multi-attribute graph partitioning problem. We then develop a set of metrics to quantify the competition and operational benefits to the airlines and propose a decision making tool for optimizing the airline alliance structure based on flight schedule data which is commonly available to the airlines and regulators through the \ac{OAG} dataset.
%
The contributions of this paper are as follows:
\begin{itemize}
	\item a metric that measures the competitiveness of a flight segment based on the alliance membership of the airlines that operate on that segment\mfabian{;}
	\item a metric that measures the market penetration capability of an airline based on the alliance membership of the airlines that operate on the routes that the airline operates on\mfabian{;}
	\item an optimization problem that maximizes the competitiveness of the routes and the market penetration capability of the airlines simultaneously.
	\item a set of algorithms to solve th\mfabian{is} optimization problem.
\end{itemize}

In Section~\ref{sec:background}, we provide background information on the data and methods used in this work. In Section~\ref{sec:optimization}, we quantify the effects on competition and airline operations in terms of market penetration and then formulate a bi-objective optimization problem using the proposed metrics. In Section~\ref{sec:algorithms}, we propose optimization algorithms for solving the problem. In Section~\ref{sec:results}, we present the results of the optimization problem. Section~\ref{sec:discussion} discusses key insights and implications of the results for competition and market consolidation. In Section~\ref{sec:conclusion}, we provide the conclusion and outline future work.

\section{Data and methods} \label{sec:background}

We first \mfabian{describe} the data available for this study and the methods used to analyze \mfabian{them}.
We obtained proprietary data from \ac{IATA} that was engineered from the \ac{OAG} dataset which includes records on passenger bookings and flight schedules \citep{OAG}.
We further \mfabian{got} information on various airlines and airports from the \ac{IATA} database.
Historical alliance memberships were \mfabian{recovered} from public sources and news articles.
The data \mfabian{are} summarized in Table~\ref{table:datasummary}.

For this study we examine the supply side of the airline industry. We use the \acp{ASM} as a measure of supply. The \acp{ASM} are a measure of the total number of seats available for sale on a flight. It is calculated by multiplying the number of seats on the aircraft by the number of miles flown. We attempt to develop decision making tools for optimizing the airline alliance structure based on flight schedule data which is commonly available to the airlines and regulators through the \ac{OAG} dataset.

The data is used to construct a multi-attribute graph of the airport network with airports as nodes $i \in \mathcal{V}$, \mkhalil{edges as the flight segments between airports $(u,v) \in \mathcal{E}$,} the airlines as the edge attribute $\tau \in \mathcal{T}$ and \acp{ASM} being the weight associated with each \mkhalil{airline} $w_\tau[u,v]$ for all $(u,v) \in \mathcal{E}$ and for all $\tau\in~\mathcal{T}$. Such an approach allows us to compactly represent the airline network and the competition between the airlines on all the routes in the network. The graph associated with each \mkhalil{airline} $\mathcal{G}_\tau$ corresponds to \mkhalil{its network}. The graph $\mathcal{G}$ is the union of all the airline graphs $\mathcal{G}_\tau$.

\begin{figure}[htbp]
	\centering
	\includegraphics[width=0.9\textwidth]{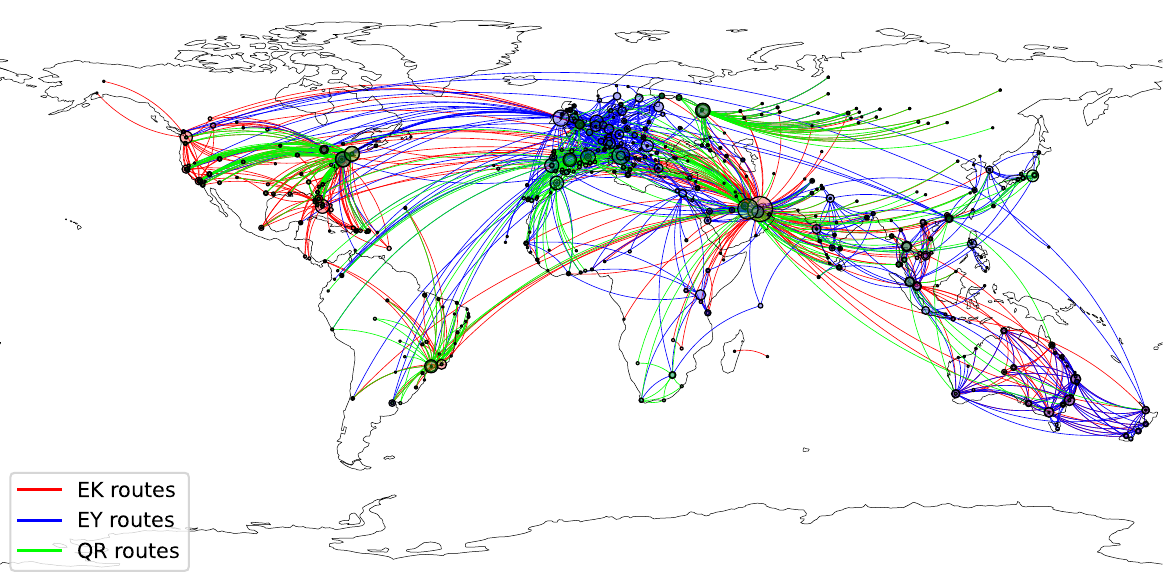}
	\caption{Graph of the \ac{IATA} network for three major middle-eastern airlines.}
	\label{fig:iatagraph}
\end{figure}

We show a subgraph of the \mkhalil{commercial aviation} network for three major middle-eastern airlines in Figure~\ref{fig:iatagraph}. The graph represents the network of airlines and the routes they operate.
%

Each airline given by an attribute $\tau$ can be assigned to an alliance $\alpha_k$ based on the alliance membership data, assuming we have $K$ alliances denoted $\alpha_1, \ldots, \alpha_K$. In the following section we define the metrics that govern the competition and the operational benefits to the airlines.

\section{Optimization problem formulation} \label{sec:optimization}



\fabian{I have removed the graph explanations as they are already given in the previous section.}

\subsection{Competition index calculation} \label{subsec:competition}

We use the multi-attribute graph of the airport network to calculate\mfabian{, for any segment $(u,v)$,} a metric $h_{u,v}$  \mfabian{representing} its competitiveness.
This metric\mfabian{,} based on the \ac{HHI} \citep{waldman2012industrial}, is estimated as 
\begin{equation} \label{eq:hhidef}
h_{u,v} = \sum_{k=1}^{K}{p\left(\alpha_k\mfabian{\,|\,(u,v)}\right)^2}\quad \forall(u,v) \in \mathcal{E},~ \mkhalil{\forall~k=1,\ldots,K},
\end{equation}
where $p(\alpha_k\mfabian{\,|\,(u,v)})$ is the conditional probability of observing airline alliance $\alpha_k$ on segment $(u,v)$. This probability can be estimated using the observed market share of alliance $\alpha_k$ on \mkhalil{segment} $(u,v)$ given by
\begin{equation} \label{eq:pcond}
p(\alpha_k \mfabian{\,|\,(u,v)}) = \dfrac{\sum_{\tau\in\mathcal{T}}x_{\tau,k} \cdot w_\tau[u,v]}{\sum_{\tau\in\mathcal{T}}{w_\tau[u,v]}},
\end{equation}
where $x_{\tau,k}$ is a binary variable that indicates whether airline $\tau$ is a member of alliance $\alpha_k$\mfabian{:}\fabian{Use the environment cases.}
\begin{equation} \label{eq:indicator}
x_{\tau,k} =
\begin{cases}
1 & \tau \in \alpha_k, \\
0 & \tau \not\in \alpha_k.
\end{cases} 
\end{equation}

We can express the ratio ${{w_\tau[u,v]}}/{\sum_{\tau\in\mathcal{T}}{w_\tau[u,v]}}$ in terms of a random variable $W$ that follows the \ac{PMF} $\mathcal{P}_{u,v}$ of the airlines on segment $(u,v)$.
We can then rewrite the conditional \mfabian{probability~\eqref{eq:pcond} in \eqref{eq:hhidef} to obtain}
\begin{equation} \label{eq:objectiveprob}
h_{u,v} = \sum_{k=1}^{K}\sum_{\tau_\in\mathcal{T}} x_{\tau,k}\cdot P\left(W=w_\tau[u,v]\right)^2 = \sum_{k=1}^{K}\sum_{\tau_\in\mathcal{T}} x_{\tau,k}\cdot \mathcal{P}_{u,v}(\tau)^2.
\end{equation}
Calculating the mean $h_{u,v}$ for all routes has a cost $\mathcal{O}\left(\lvert\mathcal{T}\rvert\cdot\lvert\mathcal{E}\rvert\right)$. Alternatively, we can use a sampling technique to estimate $p(\alpha_k\mfabian{\,|\,(u,v)})$ and reduce the cost to $\mathcal{O}\left(\lvert\mathcal{E}\rvert\right)$ for a very large number of airlines $\tau\in\mathcal{T}$.

%
\mkhalil{We construct a sample set $\mathcal{N}_{u,v}$ of airlines that are present on segment $(u,v)$ by drawing $n_\text{samples}$ independent samples from the \ac{PMF} $\mathcal{P}_{u,v}$.}
\mkhalil{We can then estimate the conditional probability $p(\alpha_k\mfabian{\,|\,(u,v)})$ as}
\begin{equation} \label{eq:pcondest}
\hat{p}(\alpha_k\mfabian{\,|\,(u,v)}) = \dfrac{1}{\lvert\mathcal{N}_{u,v}\rvert}\sum_{\mkhalil{\tau} \in \mathcal{N}_{u,v}}\mkhalil{x_{\tau,k}},
\end{equation}

We then approximate the competition index as 
\begin{equation} \label{eq:hhiestimate}
\hat{h}_{u,v} = \sum_{k=1}^{K} \left( \frac{1}{|\mathcal{N}_{u,v}|} \sum_{t \in \mathcal{N}_{u,v}} \mkhalil{x_{\tau,k}} \right)^2.
\end{equation}

\subsubsection{\Acl{MPC}} \label{subsec:marketpenetration}

\fabian{We have to discuss the rationale of the approach. We allow to take account of trips where an alliance is present on a part of the trip only, but that could cause issues for the traveller. A limit is also that there is no possibility to open new routes (where the airline is not present) in the current setting.}

We also use the multi-attribute graph for calculating \mkhalil{a} second metric that represents the \acf{MPC} of the airlines. The \ac{MPC} is a measure of the ability of an airline to access the \acp{ASM} of an alliance. We assume that the \ac{MPC} of an airline $\tau$ is proportional to the probability of observing airline $\tau$ on a route of length $L$ starting from \mkhalil{airport} $i$ \textbf{and} observing a \mkhalil{member of airline $\tau$'s} alliance $\mkhalil{\tau'} \in \alpha_k$ along the same route.
\begin{equation} \label{eq:mpcdef}
w_{\tau,i} = \sum_{k=1}^{K}x_{\tau,k} \cdot p(\tau | i, L) \cdot p(\alpha_k | i, L),
\end{equation}
These probabilities can be calculated by traversing the graph starting from \mkhalil{airport} $i$ up to a depth $L$. We first define the \ac{PMF} of \mkhalil{airport} $u$'s neighbourhood as $\mathcal{P}_u$. Sampling from the \ac{PMF} $\mathcal{P}_u$ gives us the next \mkhalil{airport} $v$. We also use the previous definition of the segment \ac{PMF} $\mathcal{P}_{u,v}$ to sample the airline $\tau$ on the segment $(u,v)$. We then calculate the \ac{MPC} of the airlines recursively as follows:
\begin{equation} \label{eq:mpcdefairline}
p(\tau|i,L) = \dfrac{1}{L}\sum_{v\in \mathcal{N}_i}{\left(\mathcal{P}_i(v) \mathcal{P}_{i,v}(\tau) + \mathcal{P}_i(v) \sum_{\tau'\in\mathcal{T}}\mathcal{P}_{i,v}(\tau')\cdot p(\tau|v,L-1) \right)},
\end{equation}
\begin{equation} \label{eq:mpcdefalliance}
p(\alpha_k|i,L) = \dfrac{1}{L}\sum_{v\in \mathcal{N}_i}{\left(\mathcal{P}_i(v) \sum_{\tau\in\mathcal{T}}x_{\tau,k}\cdot\mathcal{P}_{i,v}(\tau) + \mathcal{P}_i(v) \sum_{\tau'\in\mathcal{T}}\mathcal{P}_{i,v}(\tau')\cdot p(\alpha_k|v,L-1) \right)},
\end{equation}
where $p(\alpha_k|i,0) := 0$ and $p(\tau|i,0) := 0$ for all $i\in\mathcal{V}$ and $\tau\in\mathcal{T}$.
The recursive calculation in Equations~(\ref{eq:mpcdefairline}) and (\ref{eq:mpcdefalliance}) can be done using a dynamic programming approach and can be computationally expensive.

We can use a sampling technique to estimate the \ac{MPC}. We do this by first launching several unbiased independent random walks from every \mkhalil{airport} $u\in\mathcal{V}$ in the graph. This results in a 3 dimensional tensor 
\[ {V} \in \mathcal{V}^{\lvert\mathcal{V}\rvert\times n_\text{walks} \times L}, \]
where ${V}_{i,j,\mkhalil{l}}$ corresponds to the sampled \mkhalil{airport}s along an independent random walk $j$ of length $L$. The \mkhalil{airport} tensor ${V}_{i,j,\mkhalil{l}}$ is constructed by sampling the next \mkhalil{airport} $v_{i,j,\mkhalil{l}+1}$ from the \ac{PMF} $\mathcal{P}_{v_{i,j,\mkhalil{l}}}$ of the current \mkhalil{airport} $v_{i,j,\mkhalil{l}}$.
\[v_{i,j,\mkhalil{l}+1}\sim\mathcal{P}_{v_{i,j,\mkhalil{l}}},\]
where $\mathcal{P}_\mkhalil{v_{i,j,{l}}}$ is the probability mass function of all the current \mkhalil{segment}s incident to \mkhalil{airport} $v_{i,j,\mkhalil{l}}$. The details of the random walk implementation are given in Algorithm~\ref{alg:randomwalk}.

We can convert the tensor $V_{i,j,\mkhalil{l}}$ to other 3 dimensional tensors 
\[{T}_{i,j,\mkhalil{l}} \in \mathcal{T}^{\lvert\mathcal{V}\rvert\times n_\text{walks} \times L-1}\]
\[{W}_{i,j,\mkhalil{l}} \in \mathbb{R}^{\lvert\mathcal{V}\rvert\times n_\text{walks} \times L-1},\]
where ${T}_{i,j,\mkhalil{l}}$ corresponds to the \mkhalil{airline} $\tau \in \mathcal{T}$ along segment $({V}_{i,j,\mkhalil{l}}, {V}_{i,j,\mkhalil{l+1}})$, while ${W}_{i,j,\mkhalil{l}}$ corresponds to the weight of the \mkhalil{airline} $\tau$ along segment $w_\tau[{V}_{i,j,\mkhalil{l}}, {V}_{i,j,\mkhalil{l+1}}]$.

The \mkhalil{airline} tensor ${T}_{i,j,\mkhalil{l}}$ is calculated by sampling from the probability mass function of the \mkhalil{airline} weights
\[T_{i,j,\mkhalil{l}}\sim\mathcal{P}_{v_{i,j,\mkhalil{l}},v_{i,j,\mkhalil{l}+1}}.\]
The corresponding weight matrix is updated with the sampled \mkhalil{airline} weight $w_{T_{i,j,\mkhalil{l}}}[v_{i,j,\mkhalil{l}},v_{i,j,\mkhalil{l}+1}]$. The details of this sampler are given in Algorithm~\ref{alg:conditional}.

Using the tensors $T$ and $W$, we may compute the \ac{MPC} of the alliances by aggregating the weights along the $j$th dimension if the \mkhalil{sampled airline}s $\tau$ are all part of the same alliance $\alpha_k$. 
\begin{equation} \label{eq:allianceweight}
\hat{p}(\alpha_k|i,L) = \dfrac{1}{n_\text{walks}\cdot L}\sum_{j=1}^{n_\text{walks}}\sum_{l=1}^{L}\mkhalil{x_{T_{i,j,l},k}},
\end{equation}
We then calculate the total weight of each airline $\tau$ along the random walk $j$. This can expressed mathematically as follows:
\begin{equation} \label{eq:airlineweight}
\hat{p}(\tau|i,L) = \dfrac{1}{n_\text{walks}\cdot L}\sum_{j=1}^{n_\text{walks}}\sum_{l=1}^{L}\mathbf{1}_\tau\left(T_{i,j,l}\right),
\end{equation}
\mkhalil{where $\mathbf{1}$ is the indicator function.} We then compute the product of (\ref{eq:allianceweight}) and (\ref{eq:airlineweight}) to compute the total gain to airline $\tau$ of joining alliance $\alpha_k$ as follows:
\begin{equation} \label{eq:mpcestimate}
\hat{w}_{\tau,i} = \sum_{k=1}^{K}x_{\tau,k} \cdot \hat{p}(\alpha_k|i,L)\cdot \hat{p}(\tau|i,L),
\end{equation}

(\ref{eq:mpcestimate}) describes the relative benefit to airline $\tau$ of joining alliance $\alpha_k$ in terms of the airline's ability to access the alliance's \acp{ASM} from airport $i$.

We take the logarithm of the average of $\hat{w}_{\tau,i}$ over all the root \mkhalil{airport}s $i\in\mathcal{V}$ to estimate the \ac{MPC} of the airlines. This can be expressed as
\begin{equation} \label{eq:mpcestimatefinal}
\hat{w}_{\tau} = \log{\left(\dfrac{1}{\left|\mathcal{V}\right|}\sum_{i\in\mathcal{V}}\hat{w}_{\tau,i}\right)}.
\end{equation}
The logarithm is used to scale the \ac{MPC} of the airlines to a more manageable range since \mkhalil{we observed that the \ac{MPC} values in our dataset approximately follow a log-normal distribution.}
This places more emphasis on the airlines with lower \ac{MPC} values.

With these metrics, we can formulate an optimization problem that maximizes the competitiveness of the routes and the \ac{MPC} of the airlines. This can be formulated as a multiobjective optimization problem as follows:
\begin{equation*} \label{eq:analyticalobjectiveapprox}
\begin{aligned}
& \underset{\left\{\alpha_1,\alpha_2,\cdots,\alpha_K\right\}\in\mathcal{T}^K}{\text{maximize}}
& & -\dfrac{\beta}{\left|\mathcal{E}\right|}\sum_{\left(u,v\right) \in \mathcal{E}} \hat{h}_{u,v} + \dfrac{\gamma}{\left|\mathcal{T}\right|}\sum_{\tau\in\mathcal{T}}\hat{w}_{\tau},
\end{aligned}
\end{equation*}
The objective function can be expressed in terms of the binary variable $x_{\tau,k}$ as follows:
\begin{equation} \label{eq:analyticalestimate}
\begin{aligned}
& \underset{\left\{\alpha_1,\alpha_2,\cdots,\alpha_K\right\}\in\mathcal{T}^K}{\text{maximize}}
& & -\dfrac{\beta}{\left|\mathcal{E}\right|}\sum_{\left(u,v\right) \in \mathcal{E}}\sum_{k=1}^{K} \left( \frac{1}{|\mathcal{N}_{u,v}|} \sum_{t \in \mathcal{N}_{u,v}} \mkhalil{x_{t,k}} \right)^2 + \\
& & & \dfrac{\gamma}{\left|\mathcal{T}\right|}\sum_{\tau\in\mathcal{T}}\log{\left(\mkhalil{\dfrac{1}{\left|\mathcal{V}\right|}}\sum_{i\in\mathcal{V}} \dfrac{\hat{p}(\tau|i,L)}{n_\text{walks}\cdot L}\sum_{k=1}^{K}\mkhalil{x_{\tau,k}}\sum_{j=1}^{n_\text{walks}}\sum_{l=1}^{L}\mkhalil{x_{T_{i,j,l},k}}\right)},
\end{aligned}
\end{equation}
%
%
where $\beta \in \mathbb{R}^{+}$ and $\gamma \in \mathbb{R}^{+}$ are the weights assigned to the two objectives.

\section{Solution algorithms} \label{sec:algorithms}

In this section, we explore a variety of algorithms to solve the alliance partitioning optimization problem formulated posed in \eqref{eq:analyticalestimate}. We use two algorithms inspired from solutions to modularity maximization for community detection problems in graph theory \citep{Clauset2004,FortHric16,Aref2022}. The optimization problem~\eqref{eq:analyticalestimate} bears some resemblance to modularity maximization in the sense that we assign a graph attribute to a community. However, our proposed optimization problem has an objective that is fundamentally different to the modularity and focuses on partitioning multi-attribute graphs rather than the nodes of a single attribute graph.

\subsection{Greedy method}\label{subsec:greedy}

We utilize a greedy heuristic method for solving the optimization problem. The greedy method is a simple and efficient algorithm that iteratively selects the best possible solution at each step \citep{Clauset2004}. The greedy method is particularly useful when the optimization problem is computationally expensive.
and the objective function is non-convex. 
The greedy method is guaranteed to find a local optimum, but not necessarily the global optimum.

The naive greedy algorithm for solving the optimization problem is given in Algorithm~\ref{alg:greedy}. The algorithm initializes each \mkhalil{airline} as a separate \mkhalil{alliance} and then iteratively merges the \mkhalil{alliance}s that result in the highest increase in the objective function. The algorithm terminates when the increase in the objective function is less than a predefined threshold or when the number of \mkhalil{alliance}s is \mkhalil{equal to} 1.

\begin{algorithm}[htbp]
\caption{Greedy algorithm for optimal partitioning}
\label{alg:greedy}

\begin{algorithmic}[1]
\Procedure{GreedyPart}{$\mathcal{G} = (\mathcal{V}, \mathcal{E}, \mathcal{T})$, $T \in \mathcal{T}^{\left|\mathcal{V}\right|\times n_\text{walks} \times L} $, $\mathcal{N}_{u,v} \forall \left(u,v\right)\in\mathcal{E}$, $\beta$, $\gamma$}
\State Initialize all \mkhalil{airline}s as separate \mkhalil{alliance}s
\Statex\hspace*{\algorithmicindent}Let $\alpha_k$ represent an \mkhalil{alliance} and $\tau \in \mathcal{T}$ represent an \mkhalil{airline}
\Statex\hspace*{\algorithmicindent}Assign each \mkhalil{airline} $\tau$ to exactly one \mkhalil{alliance} $\alpha_k$ using the following constraint:
\Statex\hspace*{\algorithmicindent}\[
\sum_{k=1}^{K} x_{\tau,k} = 1, \quad \forall \tau \in \mathcal{T}
\]
\Statex\hspace*{\algorithmicindent}Compute $\hat{p}\left(\tau|i,L\right)~\forall~\tau \in \mathcal{T},~i \in \mathcal{V}$ using \eqref{eq:airlineweight}

\State Initialize iteration counter $m \gets 1$, $K = \lvert\mathcal{T}\rvert$

\State Calculate the objective $\bar{f}_0$ using \eqref{eq:analyticalestimate}

\Repeat
\For {$p$ in $\left\{1,2,\cdots,K\right\}$}
\For {$q$ in $\left\{1,2,\cdots,K\right\}$}
\State Merge \mkhalil{alliances} $\alpha_p$ and $\alpha_q$ to form a new \mkhalil{alliance} $\alpha_{K+1}$
\State Calculate the objective $\bar{f}_{(p,q)} = \beta f_{1,(p,q)} + \gamma f_{2,(p,q)}$ in \eqref{eq:analyticalestimate}
\Statex\hspace*{\algorithmicindent}\hspace*{\algorithmicindent}\hspace*{\algorithmicindent}\hspace*{\algorithmicindent}excluding the merged \mkhalil{alliances} $\alpha_k \notin \left\{p,q\right\}$
\EndFor
\EndFor
\Statex\hspace*{\algorithmicindent}\hspace*{\algorithmicindent}find the best pair $(p,q) = \underset{p,q}{\text{argmax}}f_{(p,q)}$
\Statex\hspace*{\algorithmicindent}\hspace*{\algorithmicindent}$m \gets m + 1$, $K \gets K - 1$
\Until{$f_m - f_{m-1} \leq 0$ \textbf{or} $K \mkhalil{=} 1$}
\State \textbf{return} partitioning of $\mathcal{G}$ into \mkhalil{alliances} $\left\{\alpha_1,\alpha_2,\cdots,\alpha_K\right\}$
\EndProcedure
\end{algorithmic}
\end{algorithm}

A drawback of the greedy method is that it may get stuck in a local optimum and not find the global optimum. To mitigate this, we can run the greedy algorithm multiple times with different initializations and select the partitioning that gives the highest objective value. The greedy method is computationally efficient and can be used to solve large-scale optimization problems.

Another drawback of the above algorithm is the presences of the nested loops which can lead to a high computational cost $\mathcal{{O}}(K^3)$ in the worst case where $K$ is the number of \mkhalil{alliances}. We can use an importance sampling technique to choose the most promising pair of \mkhalil{alliances} $(p,q)$ to merge at each iteration based on the expected increase in the objective. The expected increase in objective can be calculated as follows:
\begin{equation} \label{eq:expectedmod}
\mathbb{E}[\bar{f}_{(p,q)}] = \bar{f}_{p} + \bar{f}_{q},
\end{equation}
where $\bar{f}_{p}$ and $\bar{f}_{q}$ are given by (\ref{eq:analyticalestimate}). This heuristic can reduce the computational cost to $\mathcal{O}(K)$ and provide some level of exploration in the search space.

\subsection{\Ac{MIQP} formulation} \label{subsec:exact}

Solving the optimization problem using a \ac{MIP} algorithm involves formulating the problem as an \ac{ILP} model or \ac{MIQP}. Gurobi is highly capable of solving large-scale \ac{ILP} problems efficiently, but setting up the problem correctly is crucial. \ac{MIQP} problems are more complex and may require more computational resources.


The objective involves maximizing the sum of $h_{u,v}$ and $w_{\tau,i}$.
\mkhalil{The first and second terms in \eqref{eq:analyticalestimate} are a quadratic combination} of the binary variables $x_{\tau, k}$ and requires an \ac{MIQP} approach.


We note that the logarithm in \mkhalil{\eqref{eq:analyticalestimate}} cannot be directly implemented in a quadratic program. Instead, we can use a \ac{PWL} approximation of the logarithm function. This involves introducing additional auxiliary variables and constraints.

We introduce auxiliary variables $y_{\tau,i,k}$ to represent the logarithm of the sum of the \ac{MPC} of the airlines. We can then write the logarithm as a linear combination of the auxiliary variables:
\begin{equation} \label{eq:finalobjaux}
\bar{f} = \dfrac{\beta}{\left|\mathcal{E}\right|}\sum_{(u,v)\in\mathcal{E}}\sum_{k=1}^{K} \left( \frac{1}{|\mathcal{N}_{u,v}|} \sum_{t \in \mathcal{N}_{u,v}} x_{t, k} \right)^2 - \dfrac{\gamma}{\left|\mathcal{T}\right|}\sum_{\tau\in\mathcal{T}}y_{\tau},
\end{equation}
where $\varepsilon \leq y_{\tau} \leq 1$ is the auxiliary variable and $\varepsilon$ is a small positive constant.

The auxiliary constraint can be written as follows:
\begin{equation} \label{eq:auxiliary}
y_{\tau} = \log{z_{\tau}},
\end{equation}
where $z_{\tau}$ is a continuous variable representing the argument of the logarithmic function inside the objective function. i.e.,
\[
\text{let } z_{\tau} = \left(\dfrac{1}{\left|\mathcal{V}\right|}\sum_{i\in\mathcal{V}}\dfrac{\hat{p}(\tau|i,L)}{n_\text{walks}\cdot L}\sum_{k=1}^{K}x_{\tau,k}\sum_{j=1}^{n_\text{walks}}\sum_{l=1}^{L}x_{T_{i,j,l},k}\right),
\]
We approximate the constraint in (\ref{eq:auxiliary}) using a \ac{PWL} approximation. This involves introducing additional auxiliary variables and constraints to linearize the logarithm function.
\begin{equation} \label{eq:pwlapprox}
y_{\tau} = \text{PWL}(z_{\tau}, (\varepsilon, 1), (I_1, I_2, \ldots, I_{n_\text{intervals}}), (V_1, V_2, \ldots, V_{n_\text{intervals}})),
\end{equation}
where $\text{PWL}$ denotes the \ac{PWL} approximation function. Here, $(\varepsilon, 1)$ defines the domain of the approximation, $(I_1, I_2, \ldots, I_{n_\text{intervals}})$ are the breakpoints of the domain, and $(V_1, V_2, \ldots, V_{n_\text{intervals}})$ are the corresponding function values at these breakpoints. $\varepsilon$ is a small positive constant. The breakpoints and values can be calculated as follows:
\begin{equation} \label{eq:breakpoints}
    I_k = \varepsilon + \frac{\mkhalil{k-1}}{n}(1-\varepsilon), \quad \mkhalil{k=1,2,\ldots,n_\text{intervals}},
\end{equation}
\begin{equation} \label{eq:values}
    V_k = \log(I_k), \quad \mkhalil{k=1,2,\ldots,n_\text{intervals}}.
\end{equation}

\mkhalil{Furthermore, alliance membership is mutually exclusive.} Each \mkhalil{airline} $\tau$ should be in exactly one subset $\alpha_k$:
\begin{equation} \label{eq:constraint}
\sum_{k=1}^K x_{\tau, k} = 1 \, \forall \tau \in \mathcal{T}.
\end{equation}

The final optimization problem can be written as follows:
\begin{equation} \label{eq:finalopt}
\begin{aligned}
& \underset{x_{\tau, k} \in \{0, 1\}}{\text{minimize}}
& & \dfrac{\beta}{\left|\mathcal{E}\right|}\sum_{(u,v)\in\mathcal{E}}\sum_{k=1}^{K} \left( \frac{1}{|\mathcal{N}_{u,v}|} \sum_{t \in \mathcal{N}_{u,v}} x_{t, k} \right)^2 - \dfrac{\gamma}{\left|\mathcal{T}\right|}\sum_{\tau\in\mathcal{T}}y_{\tau} \\
& \text{subject to}
& & \sum_{k=1}^K x_{\tau, k} = 1 \, \forall \tau \in \mathcal{T} \\
& & & y_{\tau} = \text{PWL}(z_{\tau}, (\varepsilon, 1), (I_1, I_2, \ldots, I_{n_\text{intervals}}), (V_1, V_2, \ldots, V_{n_\text{intervals}})) \\
& & & z_{\tau} = \left(\dfrac{1}{\left|\mathcal{V}\right|}\sum_{i\in\mathcal{V}}\dfrac{\hat{p}(\tau|i,L)}{n_\text{walks}\cdot L}\sum_{k=1}^{K}x_{\tau,k}\sum_{j=1}^{n_\text{walks}}\sum_{l=1}^{L}x_{T_{i,j,l},k}\right) \\
& & & x_{\tau, k} \in \{0, 1\}, \quad y_{\tau} \in \left(-\infty,0\right], \quad z_{\tau} \in \left(0,1\right].
\end{aligned}
\end{equation}

The following section explores the effectiveness of the greedy and \ac{MIQP} algorithms in solving the optimization problem on a toy example and a real-world example in Sections~\ref{subsec:resultstoy} and \ref{subsec:resultsiata}, respectively.

\section{Solution and results} \label{sec:results}

In this section, we demonstrate the effectiveness of the greedy and \ac{MIQP} algorithms in solving the optimization problem formulated in Section~\ref{subsec:exact}. We first apply the algorithms to a toy example where the optimal solution can be computed exactly. We then apply the algorithms to a real-world example using the \ac{IATA} dataset.

\subsection{Benchmark problem} \label{subsec:resultstoy}

To create the toy example, we generate a random graph with 20 \mkhalil{airport}s and 2000 \mkhalil{segment}s. We also generate 6 \mkhalil{airline}s and randomly assign them to the \mkhalil{segment}s. We then run the greedy and \ac{MIQP} algorithms on this toy example.

\begin{figure}[H]
\centering
\includegraphics[width=0.5\textwidth]{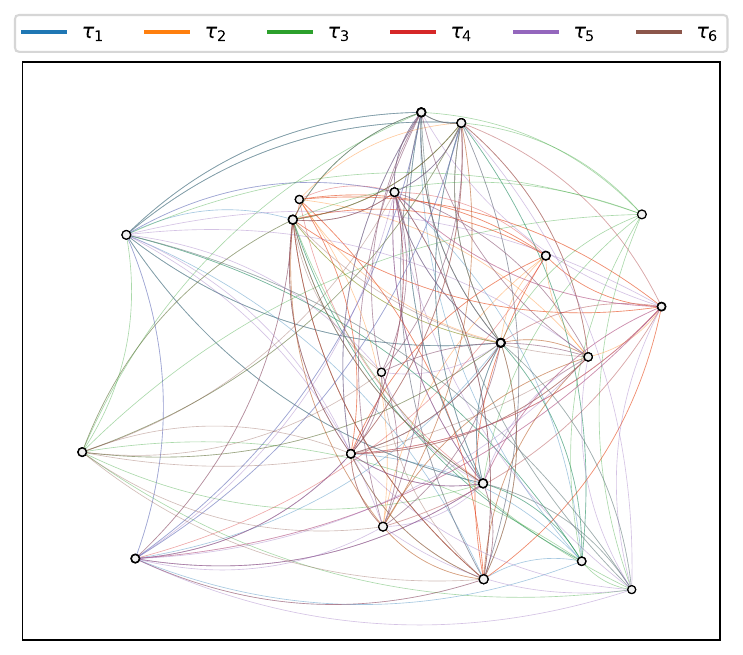}
\caption{Toy example graph}
\label{fig:toygraph}
\end{figure}

The statistics of the toy example are shown in Table~\ref{table:toystats}.
\begin{table}[H]
\centering
\caption{Toy example statistics}
\begin{tabular}{|c|c|l|}
\hline
\textbf{Parameter} & \textbf{Value} & \textbf{Description}\\
\hline
$\lvert\mathcal{V}\rvert$ & 20 & number of \mkhalil{airport}s \\
$\lvert\mathcal{E}\rvert$ & 2000 & number of \mkhalil{segment}s \\
$\lvert\mathcal{T}\rvert$ & 6 & number of \mkhalil{airline}s \\
\hline
\end{tabular}

\label{table:toystats}
\end{table}
We visualize the graph and the \mkhalil{airline}s in Figure~\ref{fig:toygraph}.

\begin{figure}[htbp]
\centering
\begin{subfigure}{0.5\textwidth}
\centering
\includegraphics[height=6.5cm]{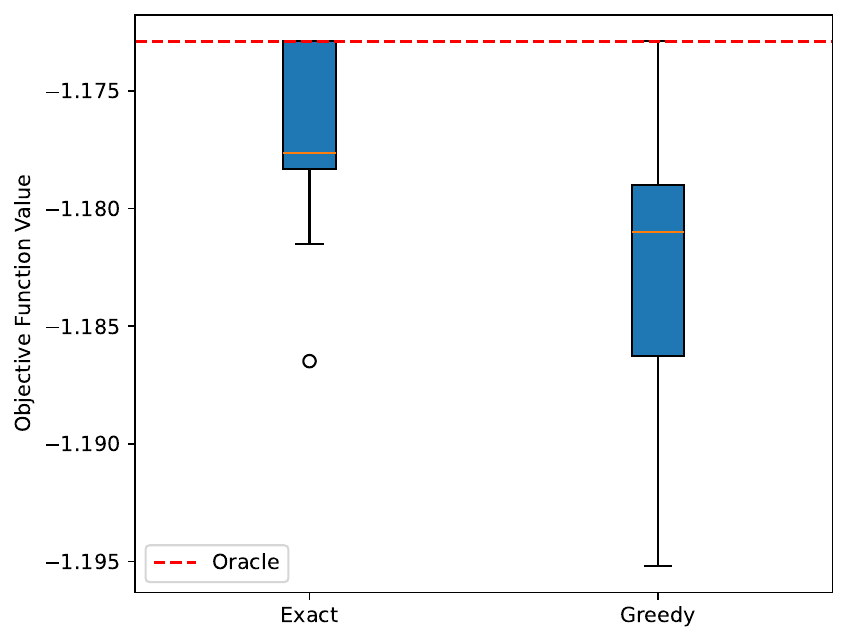}
\caption{Box plot of objective function values}
\label{fig:boxplottoy}
\end{subfigure}%
\begin{subfigure}{0.5\textwidth}
\centering
\includegraphics[height=6.5cm]{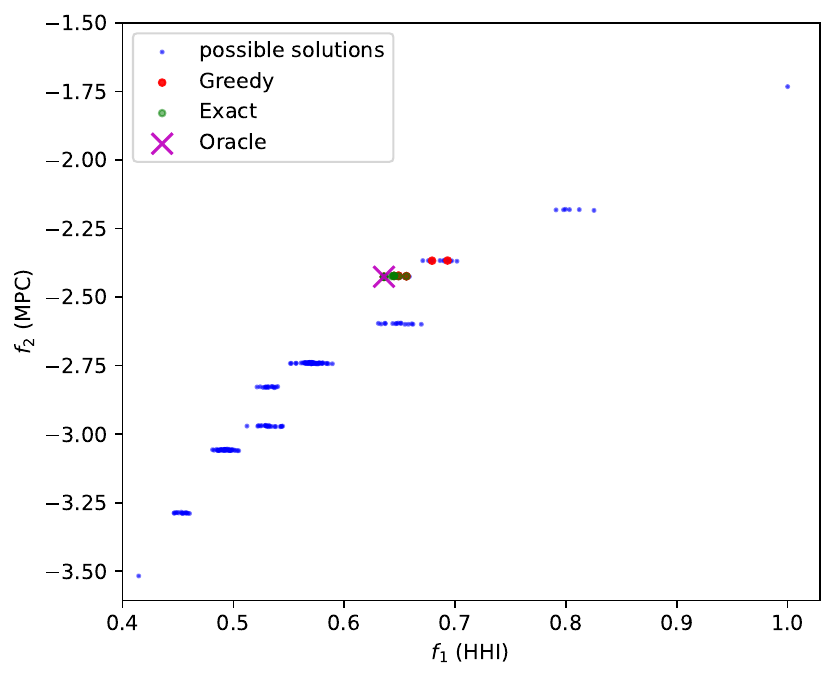}
\caption{Trade-off between \ac{HHI} and \ac{MPC}}
\label{fig:paretofrontier}
\end{subfigure}
\caption{Comparison of the \ac{MIQP} and greedy algorithms with different realizations of the random sampling. (a) Box plot of the objective function values for the exact and greedy methods. (b) Trade-off between the \ac{HHI} and \ac{MPC} in the solution space.}
\label{fig:toyproblemstats}
\end{figure}

We first solve the problem exactly by enumerating the possible partitions of the \mkhalil{airline}s into \mkhalil{alliances} and evaluating the following objective function without approximating the \ac{HHI} and \ac{MPC} terms in (\ref{eq:analyticalobjectiveexact}):
\begin{equation} \label{eq:analyticalobjectiveexact}
\begin{aligned}
& \underset{\left\{\alpha_1,\alpha_2,\cdots,\alpha_K\right\}\in\mathcal{T}^K}{\text{maximize}}
& & -\dfrac{\beta}{\left|\mathcal{E}\right|}\sum_{\left(u,v\right) \in \mathcal{E}} {h}_{u,v} + \dfrac{\gamma}{\left|\mathcal{T}\right|}\sum_{\tau\in\mathcal{T}}{w}_{\tau},
\end{aligned}
\end{equation}
The results of the entire enumeration of the solution space are shown in Figure~\ref{fig:paretofrontier}. The best solution out of all the possible enumerations is marked as the oracle solution in Figure~\ref{fig:paretofrontier} and is used as a benchmark to compare the greedy and \ac{MIQP} algorithms.

Next, we sample the graph using the random walk technique described in Section~\ref{subsec:marketpenetration} and estimate $\hat{h}_{u,v}$ and $\hat{w}_{\tau}$ using Equations~(\ref{eq:hhiestimate}) and (\ref{eq:mpcestimatefinal}) using the settings in Table~\ref{table:toyproblemsettings}.
\begin{table}[H]
\centering
\caption{Optimization problem settings}
\begin{tabular}{|c|c|l|}
\hline
\textbf{Parameter} & \textbf{Value} & \textbf{Description}\\
\hline
$\lvert\mathcal{N}_{u,v}\rvert$ & 50 & number of samples per \mkhalil{segment} \\
$n_\text{walks}$ & 50 & number of random walks launched per \mkhalil{airport} \\
$L$ & 2 & length of random walk \\
$\beta$ & 0.7 & weight of competition index objective \\
$\gamma$ & 0.3 & weight of \ac{MPC} objective \\
\hline
\end{tabular}
\label{table:toyproblemsettings}
\end{table}
We then solve the optimization problem using the greedy algorithm described in Algorithm~\ref{alg:greedy}. We set $\beta=0.7$ and $\gamma=0.3$ and run the algorithm until termination. We also solve the optimization problem using the \ac{MIQP} described in Section~\ref{subsec:exact}.

We also generate $n_\text{samples}=10$ realizations of the randomly sampled variables $t\in\mathcal{N}_{u,v} \forall (u,v)\in\mathcal{E}$ and $T_{i,j,k} \forall i\in\mathcal{V}, j\in\left\{1,2,\cdots,n_\text{walks}\right\}, k\in\left\{1,2,\cdots,L\right\}$ that are used to estimate the \ac{HHI} and \ac{MPC} terms in the objective function. We solve the optimization problem using the \ac{MIQP} and greedy algorithms for each of the 10 realizations to investigate the statistics of the estimated objective function and the sensitivity of the solution to the random sampling.

We compute the final distribution of the objective function values of the \ac{MIQP} and greedy solutions (without estimation) to compare them against the global maximum obtained by the oracle for all subsequent discussions in this section. The results of this comparison are shown in Figure~\ref{fig:boxplottoy}.

We report the numerical results of the median solution for the oracle, greedy, and \ac{MIQP} algorithms in Table~\ref{table:toyresults}.
\begin{table}[htbp]
\centering
\caption{Exact method results}
\begin{tabular}{cccc}
\hline
\textbf{Value} & \textbf{Oracle} & \textbf{Greedy (median)} & \textbf{Exact (median)} \\
\hline
$\alpha_1^*$	& $\left\{\tau_1, \tau_3, \tau_4\right\}$ 	& $\left\{\tau_1, \tau_4, \tau_6\right\}$ 	& $\emptyset$ 								\\
$\alpha_2^*$ 	& $\left\{\tau_2, \tau_5, \tau_6\right\}$ 	& $\left\{\tau_2, \tau_3, \tau_5\right\}$ 	& $\emptyset$ 								\\
$\alpha_3^*$ 	& $\emptyset$ 								& $\emptyset$ 								& $\left\{\tau_1, \tau_5, \tau_6\right\}$ 	\\
$\alpha_4^*$ 	& $\emptyset$ 								& $\emptyset$ 								& $\left\{\tau_2, \tau_3, \tau_4\right\}$ 	\\
$\alpha_5^*$ 	& $\emptyset$ 								& $\emptyset$ 								& $\emptyset$ 								\\
$\alpha_6^*$ 	& $\emptyset$ 								& $\emptyset$ 								& $\emptyset$ 								\\
Objective  		& \mkhalil{\textbf{-1.173}} 				& \mkhalil{-1.183} 							& \mkhalil{-1.176} 							\\
HHI 			& \mkhalil{-0.636} 							& \mkhalil{-0.638} 							& \mkhalil{-0.650} 							\\
MPC 			& \mkhalil{-2.426} 							& \mkhalil{-2.457} 							& \mkhalil{-2.427} 							\\
\hline
\end{tabular}
\label{table:toyresults}
\end{table}
We note that in some instances, the greedy and \ac{MIQP} algorithms are able to recover the oracle solution. The \ac{MIQP} solution has less variance in the objective function values compared to the greedy solution and has a better median objective value. 

\begin{figure}[htbp]
\centering
\begin{subfigure}{0.5\textwidth}
\centering
\includegraphics[width=\textwidth]{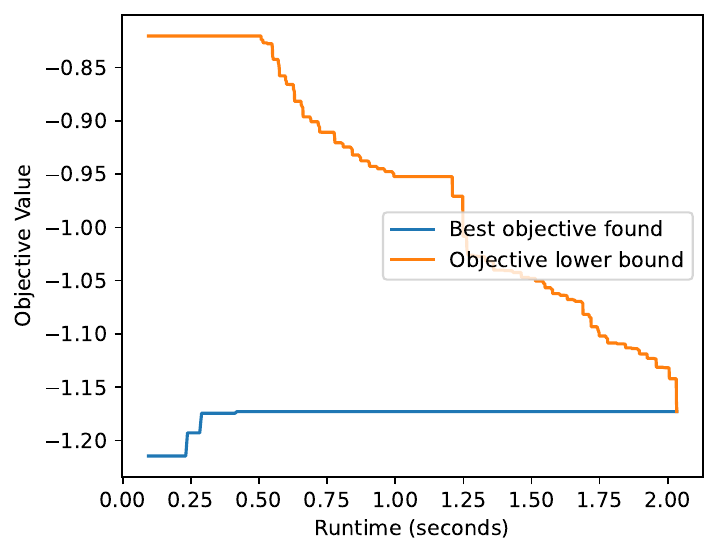}
\caption{Optimal partitioning membership}
\label{fig:exactobjtoy}
\end{subfigure}%
\begin{subfigure}{0.5\textwidth}
\centering
\includegraphics[width=\textwidth]{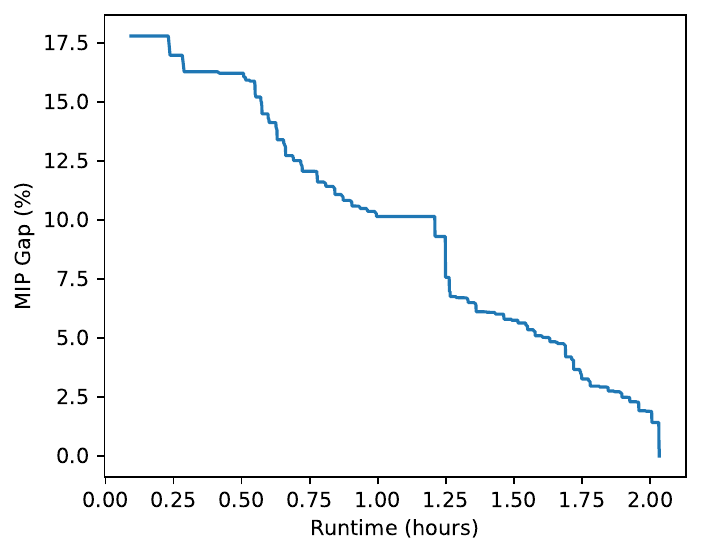}
\caption{Alliance membership}
\label{fig:extractgaptoy}
\end{subfigure}
\caption{Visualization of optimal partitioning with (a) optimal membership shown and (b) the alliance membership shown.}
\label{fig:exactconvergencetoy}
\end{figure}

We report the convergence of the \ac{MIQP} algorithm for the median solution in Figure~\ref{fig:exactconvergencetoy}. The algorithm terminates when an \ac{MIP} gap of $10^{-6}$ is reached.

\subsection{Results on \ac{IATA} network} \label{subsec:resultsiata}

As described in Section~\ref{sec:background}, we obtain data from \ac{IATA} that describes the supply and demand data along different routes in the form of \acp{ASM} and \acp{RPM}, respectively. 
%
The statistics of the \ac{IATA} network are shown in Table~\ref{table:iatastats}.
\begin{table}[H]
\centering
\caption{Toy example statistics}
\begin{tabular}{|c|c|l|}
\hline
\textbf{Parameter} & \textbf{Value} & \textbf{Description}\\
\hline
$\lvert\mathcal{V}\rvert$ & 3680 & number of \mkhalil{airport}s \\
$\lvert\mathcal{E}\rvert$ & 160732 & number of \mkhalil{segment}s \\
$\lvert\mathcal{T}\rvert$ & 580 & number of \mkhalil{airline}s \\
\hline
\end{tabular}

\label{table:iatastats}
\end{table}

We solve the optimization problem using the greedy algorithm described in Algorithm \ref{alg:greedy}. We set $\beta=0.25$ and $\gamma=0.75$ and run the algorithm until termination. Table~\ref{table:problemsettings} below summarizes the optimization problem settings.

\begin{table}[H]
\centering
\caption{Optimization problem settings}
\begin{tabular}{|c|c|l|}
\hline
\textbf{Parameter} & \textbf{Value} & \textbf{Description}\\
\hline
$\lvert\mathcal{N}_{u,v}\rvert$ & 100 & number of samples per \mkhalil{segment} \\
$n_\text{walks}$ & 20 & number of random walks launched per \mkhalil{airport} \\
$L$ & 3 & length of random walk \\
$K$ & 580 & number of alliances to be detected \\
\hline
\end{tabular}
\label{table:problemsettings}
\end{table}

\subsection{Greedy optimization results} \label{subsec:greedyresults}

The greedy algorithm terminates after $365$ iterations. This means, that 215 \mkhalil{alliances} are formed out of a total of 580 airlines. 

We construct a secondary graph where the \mkhalil{airport}s are the airlines and the edges are the codeshared \acp{ASM} between the airlines. This graph helps us visualize the airline alliances and any partitioning obtained via the algorthms in this paper in the context of codesharing.

\begin{figure}[htbp]
\centering
\begin{subfigure}{0.5\textwidth}
\centering
\includegraphics[width=\textwidth]{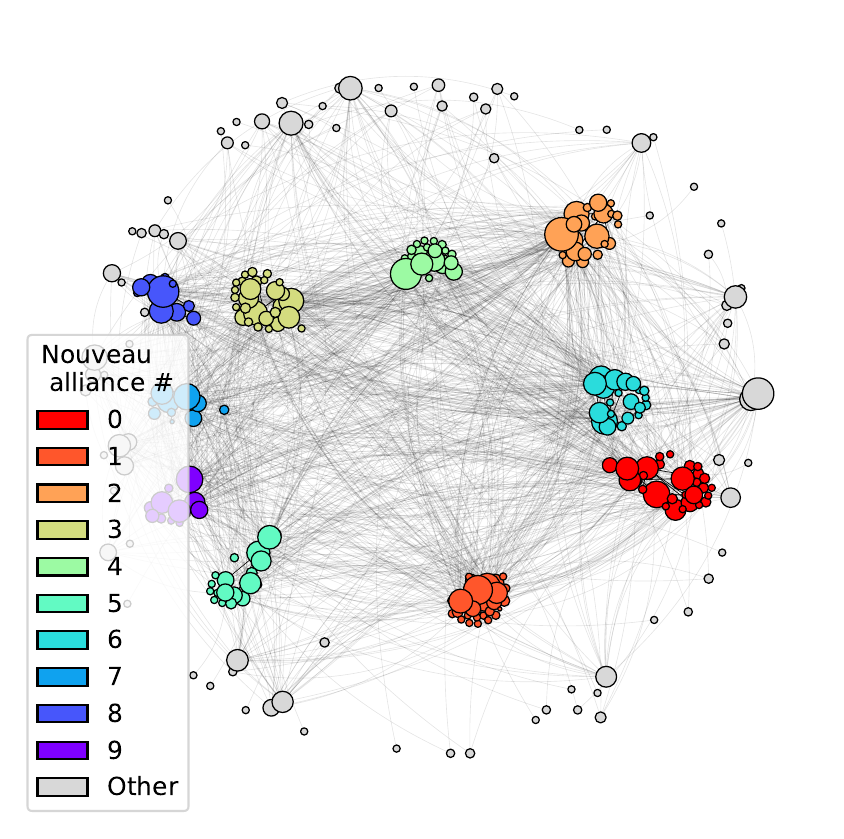}
\caption{Optimal partitioning membership}
\label{fig:partoptim}
\end{subfigure}%
\begin{subfigure}{0.5\textwidth}
\centering
\includegraphics[width=\textwidth]{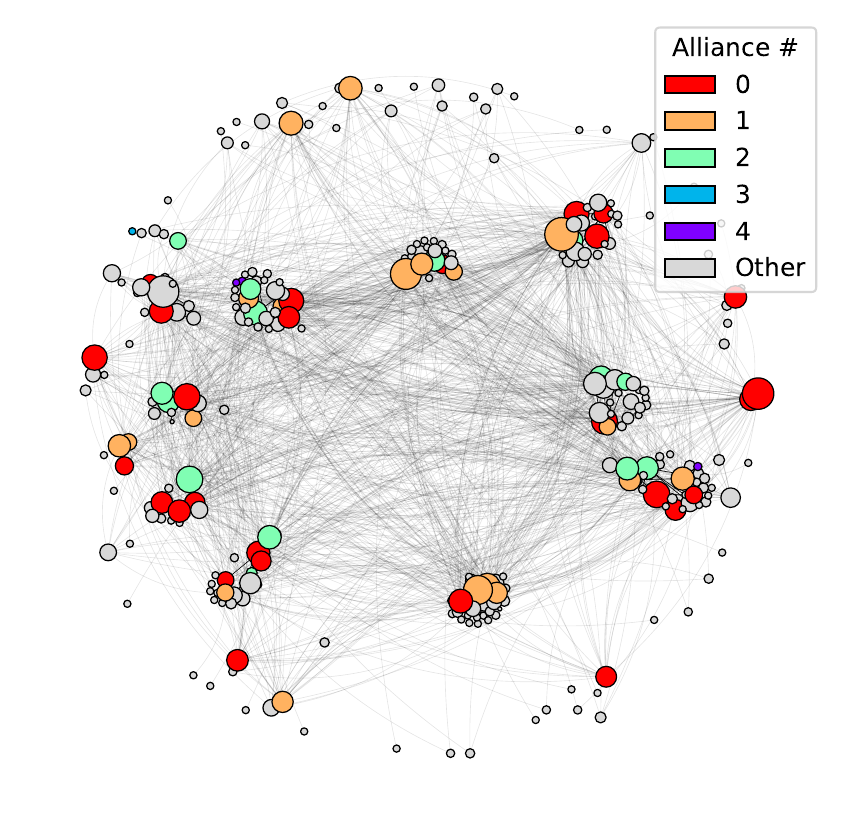}
\caption{Alliance membership}
\label{fig:partalliance}
\end{subfigure}
\caption{Visualization of optimal partitioning with (a) optimal membership shown and (b) the alliance membership shown.}
\label{fig:partition}
\end{figure}

We use the Fruchterman-Reingold algorithm to project the airlines into a 2D space and visualize the optimal partitioning in Figure~\ref{fig:partition}. We also show the alliance membership of the airlines in Figure~\ref{fig:partalliance} \citep{fruchterman1991}.

\begin{figure}[htbp]
\centering
\includegraphics[width=0.8\textwidth]{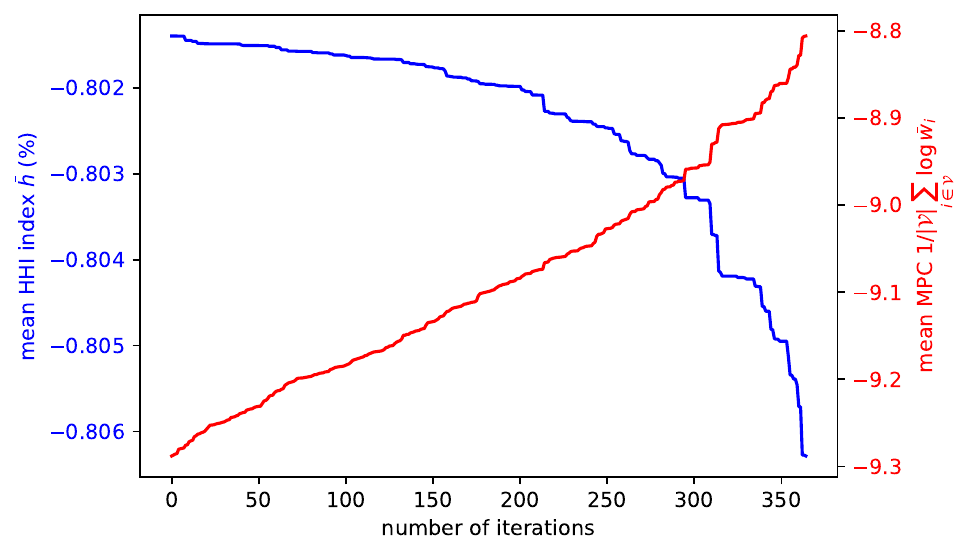}
\caption{Convergence of the greedy algorithm.}
\label{fig:convergence}
\end{figure}

The algorithmic convergence is shown in Figure~\ref{fig:convergence}.
The algorithm trades off some competitiveness to increase the \ac{MPC} of the airlines. The importance of each objective can be adjusted using the weights $\beta$ and $\gamma$. 

We compute the final objective values using an independent realization of the random walks for a fair comparison with other algorithms. 

The value of the \ac{MPC} term is $1/{\lvert\mathcal{V}\rvert\lvert\mathcal{T}\rvert}\sum_{\tau\in\mathcal{T}}\sum_{i\in\mathcal{V}}\log{\hat{w}_{\tau,i}} = -8.7856$, i.e., the mean \ac{MPC} was $e^{-8.7856}$. The value of the \ac{HHI} index is $\bar{h} = 1/{\lvert\mathcal{E}\rvert}\sum_{(u,v)\in\mathcal{E}}h_{u,v} = 0.8073$.

\subsection{Exact optimization results} \label{subsec:exactresults}

We also solve the optimization problem in (\ref{eq:finalopt}) using a \ac{MIQP} approach. We use the Gurobi solver with default settings. 

\textbf{Due to memory cost considerations}, we reduced the sampling effort to accomodate the quadratic objective terms in memory during optimization. We reduce the number of samples per \mkhalil{segment} $\lvert\mathcal{N}_{u,v}\rvert$ from 100 to 10, and the number of independent random walks per \mkhalil{airport} $n_\text{walks}$ from 20 to 10, and the number of alliances to be detected $K$ from 580 to 30.

Table~\ref{table:problemsettingsexact} below describes the problem parameter settings for the \ac{MIQP} approach.
\begin{table}[H]
\centering
\caption{Optimization problem settings for the \ac{MIQP} approach}
\begin{tabular}{|c|c|l|}
\hline
\textbf{Parameter} & \textbf{Value} & \textbf{Description}\\
\hline
$\lvert\mathcal{N}_{u,v}\rvert$ & 10 & number of samples per \mkhalil{segment} \\
$n_\text{walks}$ & 10 & number of random walks launched per \mkhalil{airport} \\
$L$ & 3 & length of random walk \\
$K$ & 30 & number of alliances to be detected \\
\hline
\end{tabular}
\label{table:problemsettingsexact}
\end{table}


Table~\ref{table:modelstats} shows the model statistics for the \ac{MIQP}.
\begin{table}[H]
\centering
\caption{Model statistics for the \ac{MIQP}}
\begin{tabular}{|l|c|}
\hline
\textbf{Statistic} & \textbf{Value}\\
\hline
Total number of binary variables & 1,080 \\
Total number of continuous variables & 17,370 \\
Total number of constraints & 579 \\
Total number of general constraints (for \ac{PWL} approximation) & 540 \\
Total number of quadratic non-zero coefficients & 2,050,380 \\
\hline
\end{tabular}
\label{table:modelstats}
\end{table}
%
%
We use the following Gurobi \ac{MIQP} settings shown in Table~\ref{table:exactsolversettings}.
\begin{table}[H]
\centering
\caption{Optimization problem settings for the \ac{MIQP} approach}
\begin{tabular}{|c|c|L{8cm}|}
\hline
\textbf{Parameter} & \textbf{Value} & \textbf{Description}\\
\hline
\texttt{TimeLimit} & 72,000 & Time limit in seconds for running the MIP solver\\
\texttt{NodefileStart} & 0.5 & When to start writing branch and bound node data to disk\\
\texttt{SoftMemLimit} & 54GB & Terminate the solver if this memory limit is reached \\
\texttt{MIPGap} & 0.001 & Terminate the solver if the gap is within 0.1\% \\
\texttt{SolutionLimit} & $\infty$ & Limits the number of feasible \ac{MIP} solutions found\\
\texttt{Heuristics} & 0.15 & Controls the agressiveness of the heuristics used\\
\texttt{MIPFocus} & 1 (feasibility) & Focus on finding new feasible solutions over proving solution optimality or improving optimality\\
\texttt{Cuts} & 0 & Disable all cuts\\
\texttt{Presolve} & 0 & Disable presolve\\
\texttt{ScaleFlag} & 1 & Enable scaling of the problem\\
\texttt{Method} & 1 & Use the dual simplex method\\
\texttt{FeasibilityTol} & 1e-2 & Feasibility tolerance for the solver\\
\texttt{IntFeasTol} & 1e-3 & Integer feasibility tolerance for the solver\\
\hline
\end{tabular}
\label{table:exactsolversettings}
\end{table}

For evaluation purposes, we compare the objective function values of the partitioning obtained via the \ac{MIQP} on the same independant realization of the random walks for a fair comparison. i.e., we compute the terms in (\ref{eq:finalopt}) using the MIQP and greedy solutions $x_{\tau,k}^\text{MIQP}$ and $x_{\tau,k}^\text{greedy}$ using an independent realization of random walks obtained with the settings in Table~\ref{table:problemsettings}.

The final objective value for the \ac{MPC} term is $1/{\lvert\mathcal{V}\rvert\lvert\mathcal{T}\rvert}\sum_{\tau\in\mathcal{T}}\sum_{i\in\mathcal{V}}\log{\hat{w}_{\tau,i}} = -8.9016$, i.e., the mean \ac{MPC} was $e^{-8.9016}$. The \ac{HHI} index value is $\bar{h} = 1/{\lvert\mathcal{E}\rvert}\sum_{(u,v)\in\mathcal{E}}\hat{h}_{u,v} = 0.8021$. 

This shows that the \ac{MIQP} solution places more emphasis on maintianing the competitiveness of the partitioning while improving the \ac{MPC} relative to the greedy optimization results in Section~\ref{subsec:greedyresults}.

\begin{figure}[htbp]
\centering
\begin{subfigure}{0.5\textwidth}
\centering
\includegraphics[width=\textwidth]{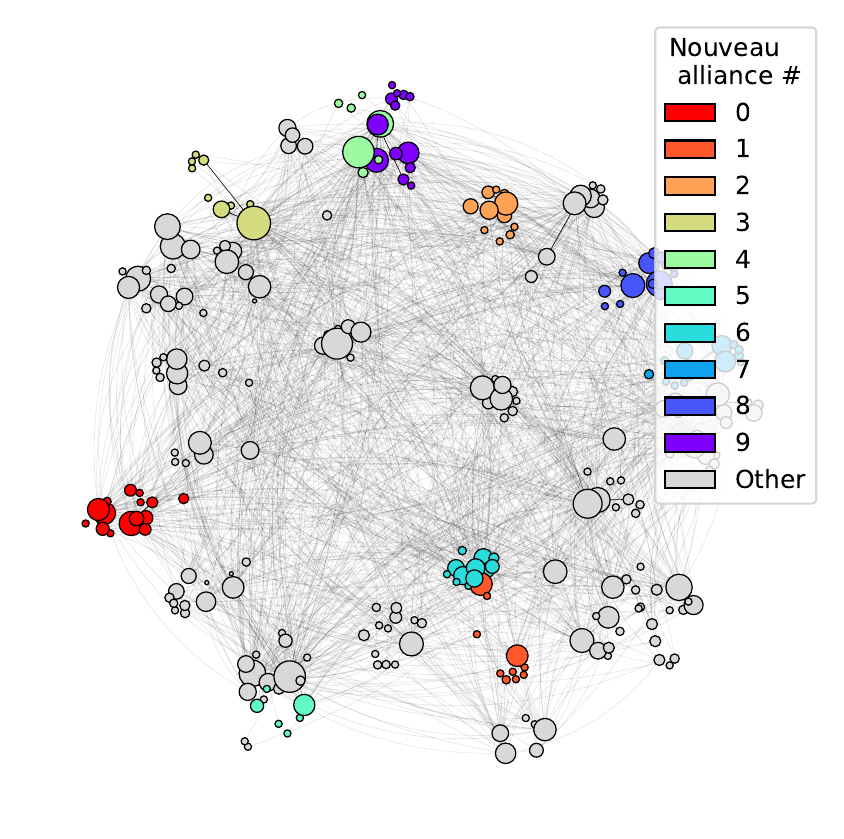}
\caption{Optimal partitioning membership}
\label{fig:partexact}
\end{subfigure}%
\begin{subfigure}{0.5\textwidth}
\centering
\includegraphics[width=\textwidth]{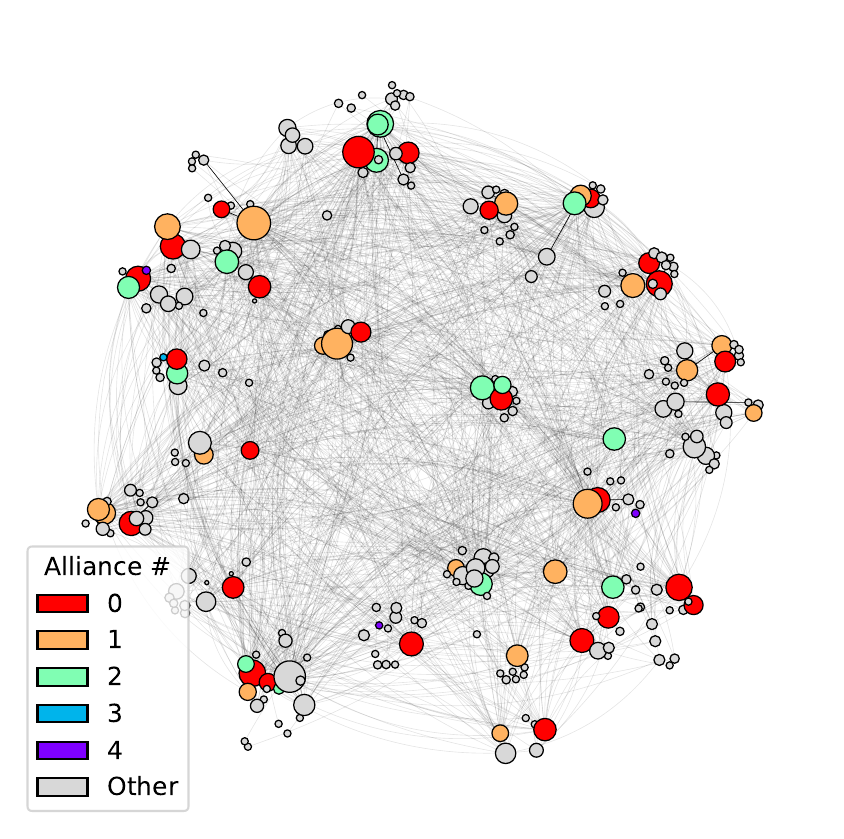}
\caption{Alliance membership}
\label{fig:partexactalliance}
\end{subfigure}
\caption{Visualization of optimal partitioning with (a) optimal membership shown and (b) the alliance membership shown.}
\label{fig:partitionexact}
\end{figure}

The airline graph is projected on 2D Cartesian space for visualization using the partitioning obtained by the \ac{MIQP} in Figure~\ref{fig:partexact}. The airline alliances are shown in Figure~\ref{fig:partexactalliance}.

Finally, we repeat all the computations performed in Sections~\ref{subsec:greedyresults} and \ref{subsec:exactresults} for $\beta=0.75$ and $\gamma=0.25$, i.e., placing more emphasis on the \ac{HHI} term in the objective function. The results are shown in Table~\ref{table:results} for comparison and are discussed in the following section.

A summary of the results of these evaluations is shown in Table~\ref{table:singleresults}

\begin{table}[h]
\centering
\caption{Results of the greedy and \ac{MIQP} optimization methods relative to the alliance partitioning}
\begin{tabular}{|l|ccc|}
\hline
\multicolumn{1}{|c|}{\centering\textbf{Method}} 	& \textbf{\ac{HHI} term} ($\%$) 	& \textbf{\ac{MPC} term} 	& \textbf{objective} $f$ 	\\ \hline
\multicolumn{4}{|c|}{\centering$\beta=0.25$, $\gamma=0.75$} 																			\\ \hline
Existing alliances 									& 80.22						& -9.2974					& -7.1736 					\\
Greedy algorithm 									& 80.73 					& \textbf{-8.7856} 			& \textbf{-6.791}			\\
\ac{MIQP} 											& \textbf{80.21} 			& -8.9016 					& -6.8767 					\\ \hline
\multicolumn{4}{|c|}{\centering$\beta=0.75$, $\gamma=0.25$} 																			\\ \hline
Existing alliances 									& 80.22						& -9.2974					& -2.926 					\\
Greedy algorithm 									& \textbf{80.16}			& -9.0497 					& -2.8636 					\\
\ac{MIQP} 											& 80.19 					& \textbf{-8.829} 			& \textbf{-2.8087} 			\\
\hline
\end{tabular}
\label{table:singleresults}
\end{table}

\begin{figure}[htbp]
\centering
\begin{subfigure}{0.5\textwidth}
\centering
\includegraphics[height=0.65\textwidth]{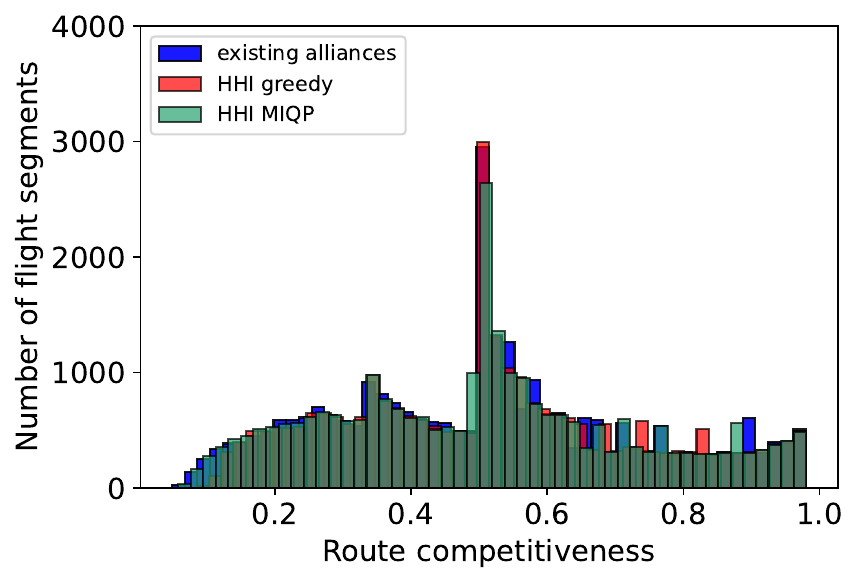}
\caption{\ac{HHI}, $\beta=0.25$, $\gamma=0.75$}
\label{fig:hhipart0250275}
\end{subfigure}%
\begin{subfigure}{0.5\textwidth}
\centering
\includegraphics[height=0.65\textwidth]{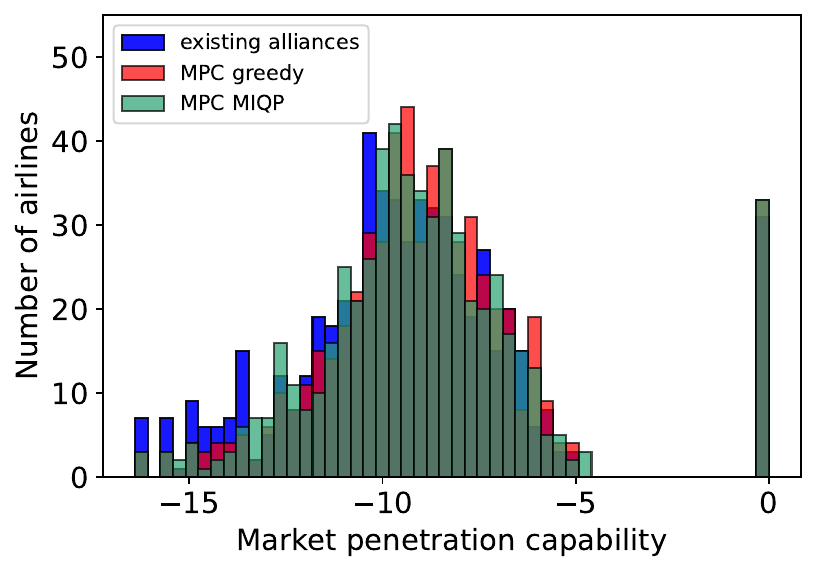}
\caption{\ac{MPC}, $\beta=0.25$, $\gamma=0.75$}
\label{fig:mpcpart025075}
\end{subfigure}
\begin{subfigure}{0.5\textwidth}
\centering
\includegraphics[height=0.65\textwidth]{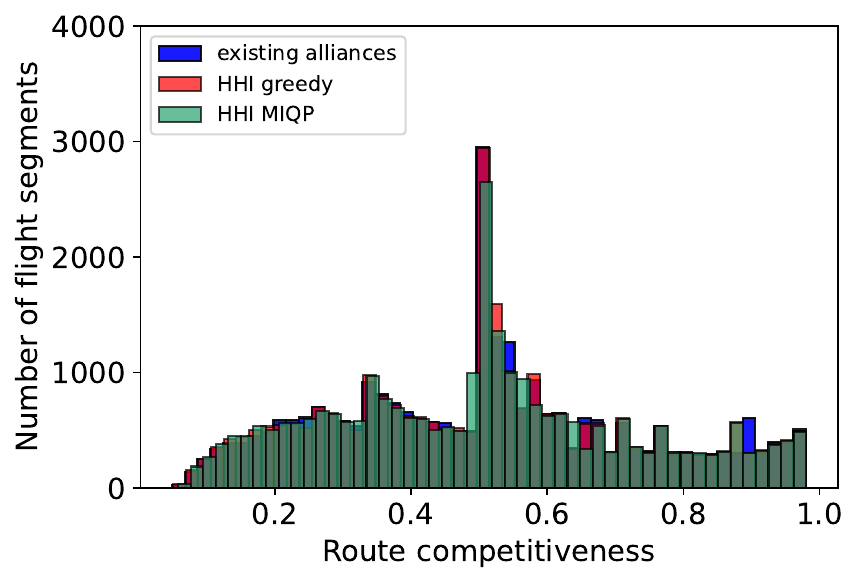}
\caption{\ac{HHI}, $\beta=0.75$, $\gamma=0.25$}
\label{fig:hhipart075025}
\end{subfigure}%
\begin{subfigure}{0.5\textwidth}
\centering
\includegraphics[height=0.65\textwidth]{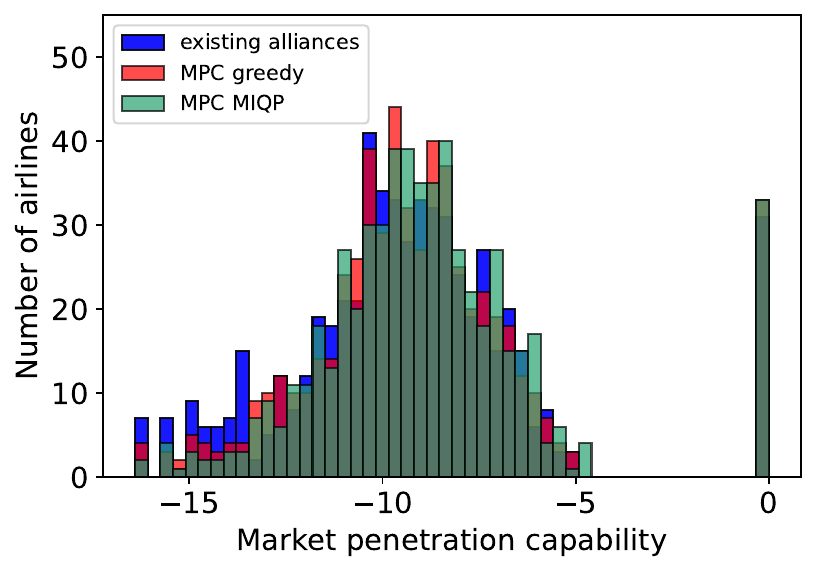}
\caption{\ac{MPC}, $\beta=0.75$, $\gamma=0.25$}
\label{fig:mpcpart075025}
\end{subfigure}
\caption{Distribution of the \ac{HHI} ($h_{u,v} \forall (u,v) \in \mathcal{E}$) and \ac{MPC} ($1/{\lvert\mathcal{V}\rvert}\sum_{i\in\mathcal{V}}\log{\bar{w}_{\tau}} \forall \tau \in \mathcal{T}, i \in \mathcal{V}$) for the greedy algorithm and the MIQP solution relative to the existing alliances.}
\label{fig:dist}
\end{figure}

In Figure~\ref{fig:dist}, we show the distribution of the competition index $h_{u,v}$ and the \ac{MPC} $\hat{w}_{\tau,i}$ for the partitioning obtained by each of the algorithms reported in Table~\ref{table:singleresults}. We show the results for both $\beta=0.25$ and $\gamma=0.75$ and $\beta=0.75$ and $\gamma=0.25$.

We note that the range of possible values is $[0, 1]$ and $[-\infty, 0]$ for the \ac{HHI} and the logarithm of \ac{MPC}, respectively. The lower the \ac{HHI}, the more competitive the segment is, while the higher the \ac{MPC}, the more efficient the airline is by virtue of its ability to access new markets.

\begin{figure}[htbp]
\centering
\begin{subfigure}{0.5\textwidth}
\centering
\includegraphics[width=\textwidth]{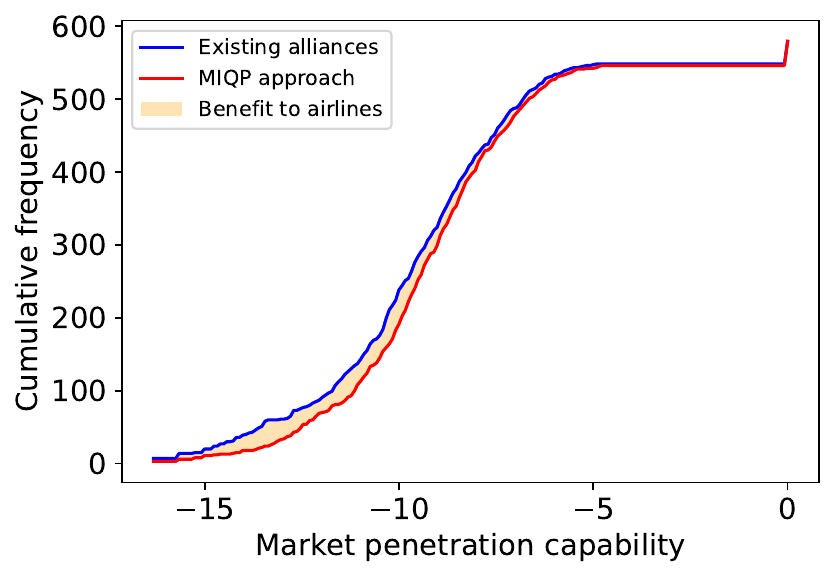}
\caption{$\beta=0.25$, $\gamma=0.75$}
\label{fig:mpccdf025075}
\end{subfigure}%
\begin{subfigure}{0.5\textwidth}
\centering
\includegraphics[width=\textwidth]{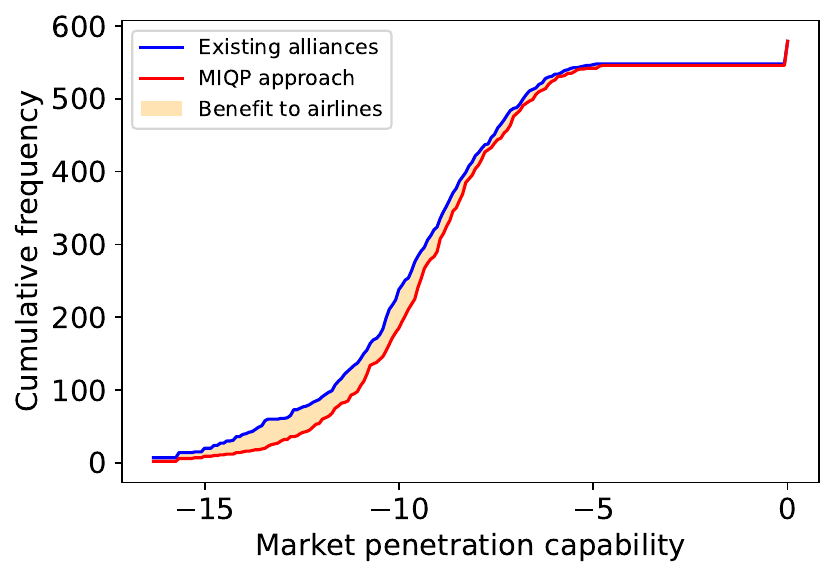}
\caption{$\beta=0.75$, $\gamma=0.25$}
\label{fig:mpccdf075025}
\end{subfigure}
\caption{Visualization the per-airline improvement in the \ac{MPC} for the \ac{MIQP} solution.}
\label{fig:mpccdf}
\end{figure}

We also examine the \ac{CMF} of the \ac{MPC} to determine the benefit of the partitioning to the airlines relative to the existing alliances. The \ac{CMF} of the \ac{MPC} is shown in Figure~\ref{fig:mpccdf}. It can be seen that the \ac{MIQP} solution improves the \ac{MPC} of the airlines when less weight is placed on the \ac{MPC} term in the objective function. The \ac{MIQP} algorithm generally performs better when the overall objective function value is small due to the smaller weight placed on the \ac{MPC} term.

To make a fair comparison between the different algorithms, 10 statistically independant optimization runs are performed with each algorithm using 10 different realizations of the random walks. The solutions are then evaluated using another 10 realizations of the random walks to estimate the objective function values. These results are discussed in detail in the following section.

\section{Discussion and comparison} \label{sec:discussion}

The following table summarizes the performance of the partitioning obtained by the existing alliances, the greedy optimization method, and the \ac{MIQP} optimization method. We conduct 10 statistically independent runs of the greedy and \ac{MIQP} optimization methods to investigate the sensitivity of the solution to the random sampling. 

\begin{figure}[htbp]
\centering
\begin{subfigure}{1\textwidth}
\centering
\includegraphics[width=0.5\textwidth]{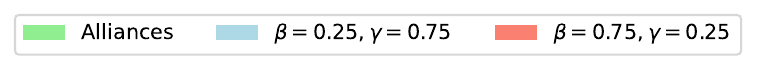}
\label{fig:comparisonlegend}
\end{subfigure}

\begin{subfigure}{0.5\textwidth}
\centering
\includegraphics[width=\textwidth]{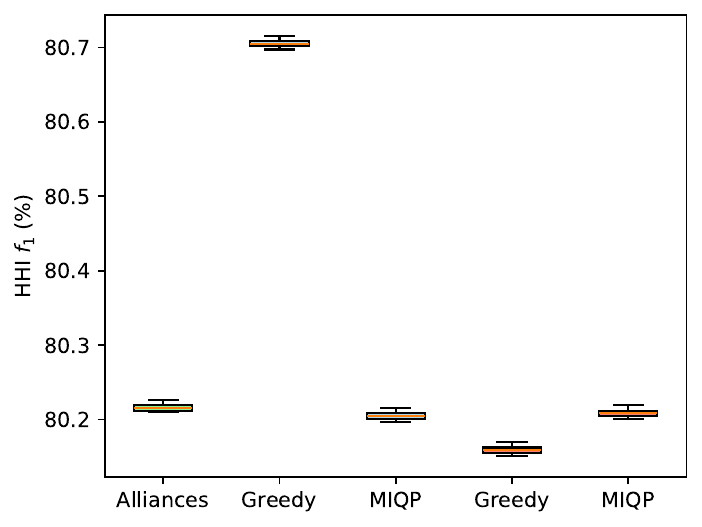}
\caption{Comparing the \ac{HHI}}
\label{fig:comparisonHHI}
\end{subfigure}%
\begin{subfigure}{0.5\textwidth}
\centering
\includegraphics[width=\textwidth]{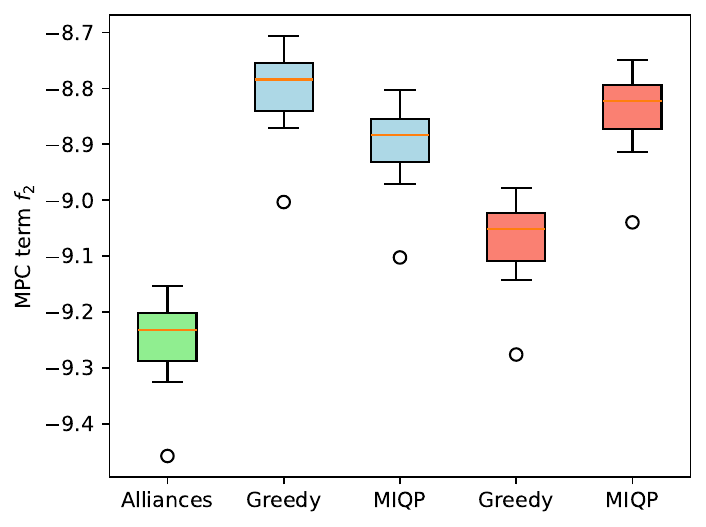}
\caption{Comparing the \ac{MPC}}
\label{fig:comparisonMPC}
\end{subfigure}

\caption{Visualization of the distribution of the estimated \ac{HHI} and \ac{MPC} for the greedy and \ac{MIQP} solutions using 10 independent runs and 10 independent realizations of the random walks.}
\label{fig:comparison}
\end{figure}

The boxplot in Figure~\ref{fig:comparison} shows the distribution of the objective function components for the greedy and \ac{MIQP} solutions. All partitions (including alliances) exhibit comparable variance values for the \ac{HHI} and \ac{MPC} terms. All algorithms (\ac{MIQP} and greedy method) result in an improvement of both objective terms relative to the alliances.The best improvement in the \ac{HHI} terms was found by the greedy method when setting the \ac{HHI} weight to $\beta=0.75$.

\begin{table}[h]
\centering
\caption{Results of the greedy and \ac{MIQP} optimization algorithms relative to the alliance partitioning}
\begin{tabular}{|l|cc|cc|cc|}
\hline
\multicolumn{1}{|c|}{\centering\textbf{Method}} 	& \multicolumn{2}{c|}{\textbf{\ac{HHI}} ($\%$)}	& \multicolumn{2}{c|}{\textbf{$\log$\ac{MPC}}} 	& \multicolumn{2}{c|}{\textbf{objective} $f$} 	\\ \hline
~ 													& mean & std.$\sigma$ 							& mean & std.$\sigma$ 							& mean & std.$\sigma$ 							\\ \hline
\multicolumn{7}{|c|}{\centering$\beta=0.25$, $\gamma=0.75$} 																																		\\ \hline
Existing alliances  								& 80.216 & 5e-05								& -9.25616 & 0.08296							& -7.14266 & 0.06222							\\
Greedy algorithm 									& 80.705 & 5e-05								& \textbf{-8.80688} & 0.08082					& \textbf{-6.80692} & 0.06061					\\
\ac{MIQP} 											& \textbf{80.205} & 6e-05						& -8.90454 & 0.08038							& -6.87892 & 0.06028							\\ \hline
\multicolumn{7}{|c|}{\centering$\beta=0.75$, $\gamma=0.25$} 																																		\\ \hline
Existing alliances 									& 80.216 & 5e-05								& -9.25616 & 0.08296							& -2.91566 & 0.02073							\\
Greedy algorithm 									& \textbf{80.159} & 5e-05						& -9.07647 & 0.08151							& -2.87031 & 0.02036							\\
\ac{MIQP} 											& 80.208 & 6e-05								& \textbf{-8.84455} & 0.07989					& \textbf{-2.8127}  & 0.01996					\\
\hline
\end{tabular}
\label{table:results}
\end{table}

We can see that both algorithms yield an improvement in the \ac{MPC} and \ac{HHI} relative to the existing alliances. The \ac{MIQP} solution is less sensitive to changes in the weighting coefficients $\beta$ and $\gamma$ as compared to the greedy method. This could be due to the large \ac{MIP} gaps encountered during the solution process and insufficient exploration of the branch and bound tree.

However, the \ac{MIQP} solution performed reliably and consistently on smaller graphs such as those in Section~\ref{subsec:resultstoy} suggesting that such approaches are more suitable for smaller graphs. The greedy algorithm is more suitable for larger graphs due to its lower memory requirements and faster convergence. This observation is consistent with the results and observations in the context of community detection and modularity maximization \citep{Aref2023}.

\section{Conclusion} \label{sec:conclusion}

In this paper, we proposed a framework for optimizing airline alliance partnerships while balancing negative externalities such as market concentration and reduced competition with potential improvements to the airlines' operational efficiency in terms of market penetration. We formulated the problem as a multi-objective optimization problem and proposed a greedy algorithm and a \ac{MIQP} approach to solve the problem. We evaluated the performance of the algorithms on a toy example and a real-world airline network obtained from \ac{IATA} data.

Our approaches yielded promising improvements to the global aviation markets compared to the existing alliances as baselines. This structured mathematical optimization approach was able to find a good balance between the network's \ac{HHI} and \acf{MPC} for a given alliance structure.

This nuanced approach can highlight the need for innovative alliance strategies and/or regulatory adjustments to ensure that the benefits of alliances are realized both for the airlines and their passengers, fostering a competitive but also cooperative aviation market \citep{Pitfield2007}. 


\section*{Acknowledgements}

\mfabian{The authors wish to acknowledge the support of the National Science and Engineering Research Council of Canada (NSERC) through the Postdoctoral Fellowship program and the Discovery Grants program. The authors also wish to thank IATA for providing the data used in this study.}



\bibliography{bibliography}
\bibliographystyle{elsarticle-harv}

\onecolumn
\pagebreak
\begin{center}
\textbf{\large Supplemental Material: Background}
\end{center}
\setcounter{equation}{0}
\setcounter{figure}{0}
\setcounter{table}{0}
\setcounter{section}{0}
\setcounter{page}{1}
\makeatletter
\renewcommand{\theequation}{S.\arabic{equation}}
\renewcommand{\thefigure}{S.\Roman{figure}}
\renewcommand{\thetable}{S.\Roman{table}}
\renewcommand{\thesection}{S.\Roman{section}}

\section{Details of the air travel dataset} \label{sec:data}

Table~\ref{table:datasummary} provides a summary of the data used for the purposes of this research. It is derived from \ac{OAG} and complemented by data provided by \ac{IATA}.

\begin{table*}[h!]
	\centering
	\renewcommand{\arraystretch}{1.5}
	\small\addtolength{\tabcolsep}{-5pt}
	\caption{Summary of available data from \ac{IPSW}}
	\label{table:datasummary}		
	\begin{tabular}{|L{5cm}|C{2cm}|C{2cm}|C{2cm}|C{2cm}|C{2cm}|}
	\hline\hline
	\mrc{3}{7cm}{\bf\centering Data Field}			& \mrc{3}{2cm}{\bf\centering Dataset}			& \multicolumn{4}{c|}{\bf sources} \\ \cline{3-6}
													& 												& \ac{OAG} 		& proprietary (\ac{IATA})	& Public 			& Derived	\\ \hline
	Name (\acs{IATA}, \acs{ICAO} designation)		& \mrc{3}{2cm}{\centering Airport}				& \cmark		& 							& 					& 			\\
	latitude/longitude								& 												& \cmark		& 							& 					& 			\\
	Region											& 												& \cmark		& 							& 					& 			\\ \hline\hline
	Name (\acs{IATA}, \acs{ICAO} designation)		& \mrc{3}{2cm}{\centering Airline}				& \cmark		& 							& 					& 			\\ \cline{3-6}
	Startup/Shutdown/merger dates					& 												& \cmark		& 							& 					& 			\\ \cline{3-6}
	Region											& 												& \cmark		& 							& 					& 			\\ \hline\hline
	Airline (\acs{IATA}, \acs{ICAO} designation)	& IOSA											&				& \cmark					& 					& 			\\ \cline{3-6}
	Date of Audit closure							& registry										&				& \cmark					& 					& 			\\ \cline{3-6}
	Registration expiry								& 												&				& \cmark					& 					& 			\\ \hline\hline
	Demand (\acs{RPM}) 								& \mrc{8}{2cm}{\centering Passenger}			& \cmark		& 							& 					& 			\\ \cline{3-6}
	Reported + Est. Pax								& 		 										& \cmark		& 							& 					& 			\\ \cline{3-6}
	Operating/Marketing flight numbers				& 		 										& \cmark		& 							& 					& 			\\ \cline{3-6}
	Flight number 									& 							 					& \cmark		& 							& 					& 			\\ \cline{3-6}
	Origin/Destinationt code						& 							 					& \cmark		& 							& 					& 			\\ \cline{3-6}
	International/Domestic flight			 		& 		 										&				&							&					& \cmark	\\ \cline{3-6}\hline\hline
	Flight numbers 									& \mrc{8}{2cm}{\centering Schedules}			& \cmark		& 							& 					& 			\\ \cline{3-6}
	Operating/Marketing flight						& 												& \cmark		& 							& 					& 			\\ \cline{3-6}
	Origin/Destination code							& 							 					& \cmark		& 							& 					& 			\\ \cline{3-6}
	Number of stops									& 							 					& \cmark		& 							& 					& 			\\ \cline{3-6}
	Stopping airports								& 							 					& \cmark		& 							& 					& 			\\ \cline{3-6}
	Supply (\acs{ASM})								& 												& \cmark		& 							& 					& 			\\ \cline{3-6}
	International/Domestic flight			 		& 		 										& 				& 							& 					& \cmark 	\\ \cline{3-6}\hline\hline
	COVID years (March 2020 - May 2021)				& \mrc{2}{2cm}{\centering misc.} 				& 				& \cmark					& \cmark			& 			\\ \cline{3-6}
	Airline alliance membership						& 												& 				& \cmark					& \cmark			& 			\\ \cline{3-6}
	\hline\hline
	\end{tabular}
\end{table*}

\section{Random walk implementation} \label{sec:randomwalker}

We use random walks to sample flight paths on the airport graph $\mathcal{G}(\mathcal{V},\mathcal{E})$. We begin by launching multiple independent random walks of length $L$ for every $i \in \mathcal{V}$ in our graph. The \mkhalil{airport} $i$ is denoted as the root \mkhalil{airport} for the random walk. During every step, the random walker checks the weights $\text{weight}(v_i,v_j)$ of the \mkhalil{segments} associated with the current \mkhalil{airport} $v_i$, where $i = \left[1,2,\cdots,L-1\right]$ and $v_j \in \mathcal{N}(v_i)$ is a neighboring \mkhalil{airport}.

The tensors of the random walk described in Section~\ref{subsec:marketpenetration} are populated by the random walker.
\begin{equation}
	{V} \in \mathcal{V}^{\lvert\mathcal{V}\rvert\times n_\text{walks} \times L},
\end{equation}
where ${V}_{i,j,k}$ corresponds to the sampled \mkhalil{airports} along an independent random walk $j$ of length $L$.

The implementation of the random walker is given by Algorithm~\ref{alg:randomwalk}.

\mkhalil{
	\begin{algorithm}
	\caption{Weighted Random Walks (from all sources)}
	\label{alg:randomwalk}
	\begin{algorithmic}[1]
		\Procedure{WeightedRandomWalks}{$\mathcal{P}_{i}~\forall~i\in\mathcal{V},\, L,\, n_{\text{walks}}$}
		\For{\textbf{each} $i \in \mathcal{V}$} 
			\For{$j = 1$ \textbf{to} $n_{\text{walks}}$}
			\State $V_{i,j,1} \gets i$
			\For{$k = 1$ \textbf{to} $L$}
				\State \textbf{sample} $V_{i,j,k+1} \sim \mathcal{P}_{V_{i,j,k}}(\cdot)$ \Comment{neighbor PMF at current \mkhalil{airport}}
			\EndFor
			\EndFor
		\EndFor
		\State \textbf{return} $V$
		\EndProcedure
	\end{algorithmic}
\end{algorithm}
}

A conditional samples is used to populate the tensors of the \mkhalil{airlines} and weights of the \mkhalil{segments} traversed by the random walker in Algorithm~\ref{alg:randomwalk}.
\begin{align}
	{T}_{i,j,k} &\in \mathcal{T}^{\lvert\mathcal{V}\rvert\times n_\text{walks} \times L-1}\\
	{W}_{i,j,k} &\in \mathbb{R}^{\lvert\mathcal{V}\rvert\times n_\text{walks} \times L-1},
\end{align}
where, ${T}_{i,j,k}$ and ${W}_{i,j,k}$ correspond to the sampled \mkhalil{airlines} $\tau \in \mathcal{T}$ and weight $w \in \mathbb{R}$, respectively, at each step of the random walk. The conditional sampler is given by Algorithm~\ref{alg:conditional}.

\mkhalil{
\begin{algorithm}
\caption{Conditional Sampling of Edge Attributes and Weights}
\label{alg:conditional}
\begin{algorithmic}[1]
\Procedure{ConditionalSampling}{$\mathcal{P}_{u,v}~\forall~u,v\in\mathcal{E},\, V,\, L$}
  \For{\textbf{each} $i \in \mathcal{V}$}
    \For{$j = 1$ \textbf{to} $n_{\text{walks}}$}
      \For{$k = 1$ \textbf{to} $L-1$}
        \State $u \gets V_{i,j,k}$,\quad $v \gets V_{i,j,k+1}$
        \State \textbf{sample} $T_{i,j,k} \sim \mathcal{P}_{u,v}(\cdot)$ \Comment{airline on segment $(u,v)$}
        \State $W_{i,j,k} \gets w_{\,T_{i,j,k}}[u,v]$ \Comment{\mkhalil{airline}-specific \mkhalil{segment} weight}
      \EndFor
    \EndFor
  \EndFor
  \State \textbf{return} $T,\,W$
\EndProcedure
\end{algorithmic}
\end{algorithm}
}

We can easily parallelize the to outer loops $i$ and $j$ in Algorithm~\ref{alg:randomwalk} and $i$, $j$, $k$ to accelerate the computation of the tensors $V$, $T$, and $W$.

\end{document}